\documentclass[11pt]{article}
\usepackage[utf8]{inputenc}
\usepackage[english]{babel}
\usepackage{amsthm}

\usepackage{epsfig}
\usepackage{amsmath}
\usepackage{amssymb}
\usepackage{lscape}
\usepackage{bm}

\usepackage{indentfirst}

\usepackage{color}

\makeatletter
\@addtoreset{equation}{section}

\makeatother

\begin{document}

\title{
\vbox{
\baselineskip 14pt
\hfill \hbox{\normalsize KEK-TH-2152
}} \vskip 1cm
\bf \Large  Dynamics of Revolving D-Branes at Short Distances
\vskip .5cm
}

\author{
Satoshi Iso$^{a,b}$  \thanks{E-mail: \tt iso(at)post.kek.jp}, 
Noriaki Kitazawa$^{c}$ \thanks{E-mail: \tt noriaki.kitazawa(at)tmu.ac.jp},
Hikaru Ohta$^{a,b}$ \thanks{E-mail: \tt hohta(at)post.kek.jp}, 
Takao Suyama$^{a}$ \thanks{E-mail: \tt tsuyama(at)post.kek.jp}
\bigskip\\
\it \normalsize
$^a$ Theory Center, High Energy Accelerator Research Organization (KEK), \\
\it  \normalsize 
$^b$Graduate University for Advanced Studies (SOKENDAI),\\
\it \normalsize Tsukuba, Ibaraki 305-0801, Japan \\
\\
\it  \normalsize 
$^c$Department of Physics, Tokyo Metropolitan University,\\
\it  \normalsize Hachioji, Tokyo 192-0397, Japan \\
\smallskip
}
\date{\today}

\maketitle

\abstract{\normalsize
We study the behavior of the effective potential between revolving D$p$-branes
 at all ranges of the distance $r$, interpolating  $r \gg l_s$ and $r \ll l_s$ ($l_s$ is the string length).
Since the one-loop open string amplitude cannot be calculated exactly,  
 we instead employ an efficient method of {\it partial modular transformation}.
The method is to perform the modular transformation partially in the moduli parameter
 and rewrite the amplitude into a sum of contributions from both of the open and closed string massless modes. 
It is nevertheless free from the double counting
 and can approximate the open string amplitudes with less than 3\% accuracy. 
From the D-brane effective field theory point of view, 
 this amounts to calculating the one-loop threshold corrections of infinitely many open string massive modes. 
We show that
 threshold corrections to the $\omega^2 r^2$ term
 of the moduli field $r$ cancel among them,
 where $\omega$ is the angular frequency of the revolution and sets the scale of supersymmetry breaking.
This cancellation suggests a possibility
 to solve the hierarchy problem of the Higgs mass in high scale supersymmetry breaking models.
}

\newpage

\section{Introduction}

D-branes in superstring theory play various essential roles,
 not only theoretically but also phenomenologically in particle physics and cosmology,
 and have been intensively and extensively investigated (see for example \cite{Blumenhagen:2006ci,Ibanez:2012zz,Baumann:2014nda}
 for phenomenological works and \cite{Dvali:1998pa,Kehagias:1999vr,Silverstein:2003hf,Easson:2007dh}
 for cosmological works).
But the dynamics has not yet been fully understood. 
Suppose two D$p$-branes are set in a target space-time. 
If they are at rest in parallel, it is a BPS configuration and stable.
When they move at a relative velocity $v$,
 very weak attractive force is induced \cite{Bachas:1995kx,Lifschytz:1996iq}.
Furthermore,
 due to the parametric resonance associated with the open string modes
 connecting D$p$-branes \cite{Bachas:1995kx,McAllister:2004gd} 
 and also due to the closed string emission \cite{AbouZeid:1999fs,Bachlechner:2013fja}, 
 the configuration loses its energy.
What is the fate of these D$p$-branes?  
They may be either separated apart or may be attracted to combine into a stack 
of D$p$-branes with an enhanced gauge symmetry \cite{Kofman:2004yc,Enomoto:2013mla}: {\it Beauty is attractive.} 
If an appropriate initial condition is given, 
they may start revolving around each other and form a bound state. 
The motivation of the present work is to investigate such a possibility.

The  mechanism we search for is similar to the Coleman-Weinberg mechanism in the following sense.
For a revolving motion, there is repulsive centrifugal potential. Thus, if there are no other attractive forces,
revolving motion cannot be a solution. 
When two D-branes relatively move, the configuration generally violates the BPS condition and 
attractive force arises radiatively. Thus the question is whether the classical centrifugal potential can be
balanced by the attractive force generated by one-loop radiative corrections of 
massive open string modes stretched between revolving D-branes. 

In order to answer whether such a stationary state exists or not,
 we calculate one-loop corrections to the interaction between two D3-branes revolving with each other. 
At large distances $r>l_s$,
 we cannot expect a bound (resonant) state,
 since the induced attractive force is too weak compared to the centrifugal repulsive force. 
Thus we focus on the behavior of the potential at shorter distances $r \lesssim l_s$. 
At short distances,
 the closed string picture is no longer valid and replaced by its dual open string picture
 \cite{Douglas:1996yp}.
Then the open string massless modes dominantly contribute to the potential between D-branes
 whose effective action is given by the supersymmetric Yang-Mills (SYM) theory.  
We first calculate the one-loop Coleman-Weinberg potential
 in a background  corresponding to the two D3-branes revolving with each other
 with the radius $r$ and the angular frequency $\omega$.
The $U(2)$ gauge symmetry in D3-brane worldvolume is spontaneously broken to $U(1) \times U(1)$ by the Higgs mechanism, where the vacuum expectation value is given by 
the diameter $2r$.\footnote{Here, we set the string scale $2\pi \alpha'=1$
 and the distance $r$ has dimension of mass.}
Open strings stretched between D3-branes acquire mass $\sim 2r$  due to the Higgs mechanism. 
In addition,
 the revolution with the angular frequency $\omega$ breaks supersymmetry
 and the masses are split by an amount of $\omega$ between bosons and fermions.  
The one-loop Coleman-Weinberg potentials  $m(r,\omega)^4 \log m(r,\omega)^2$  
generate the effective potential for the moduli field $r$.
Since the potential must vanish at $\omega=0$,
 we expect that the potential between revolving D3-branes 
 is given by $V \propto \omega^2 r^2$ in the leading order of $r$ and $\omega$ expansions. 
Therefore,
 the moduli field $r$ is expected to acquire mass proportional to the supersymmetry 
 breaking scale $\omega$.   

This is, however, not the end of the story 
 since we have infinitely many massive open string modes whose masses $M$
 are dominantly given by the string scale $m_s:= \sqrt{1/2\pi\alpha'}$,
 with additional $r$ and $\omega$ corrections; $M=M(m_s, r, \omega)$.
One-loop corrections of the massive modes to the potential of $r$ are given by
 the Coleman-Weinberg form $M^4 \log M^2$. 
Due to the supersymmetry at $\omega=0$, $m_s^2 r^2$ terms must vanish. 
Thus, expanding $M(m_s,r,\omega)$ with respect to $r$ and $\omega$,
 one-loop threshold corrections from these massive open string modes are expected 
 to become $\omega^2 r^2$. 
Since there are infinitely many massive states,
 the coefficients might be large compared to the contributions from the massless modes.
One of the motivations of the present paper is to calculate the string threshold corrections
 to the moduli field $r$ in the perspective of the hierarchy problem of the Higgs boson mass.

In this paper,
 we propose an efficient method to calculate the threshold corrections of open string massive modes
even when the open string spectrum cannot be explicitly obtained. 
The method is to combine the SYM and the supergravity calculations with appropriate cutoffs
in the moduli parameters.     
It was first suggested in \cite{Douglas:1996yp},
where a partial modular transformation
 (open-closed string duality transformation) was utilized. 
The partial modular transformation converts the ultraviolet (UV) region of the open string one-loop amplitude
 to the infrared (IR) region of the closed string tree propagations, 
 and an introduction of cutoffs in the moduli parameter avoids the double counting
 of summing both open and closed string channels. 
We apply the method to calculate the interaction potential between revolving D3-branes in parallel.

The paper is organized as follows.
In section \ref{approximation},
 we first introduce a method to efficiently calculate string threshold corrections 
 from massive open string modes in D-brane models. 
In a toy example where the open string spectrum can be exactly obtained,
we check that the method gives a very accurate approximation to the potential. 
We then apply it to a system of revolving D$p$-branes.  
In section \ref{revolving Dp},
 the contributions from massless open string modes
 are calculated in the SYM theory with a stationary revolving background. 
In section \ref{sugra},
 we calculate the contributions from massive open string modes using the supergravity theory 
 with an appropriate Schwinger parameter cutoff. 
In section \ref{Epotential},  
 we sum up these two contributions in sections \ref{revolving Dp} and \ref{sugra}. 
We explicitly evaluate the effective potential at short distances $r \ll m_s$
 by expanding the formulae derived in the previous sections, and draw the shape of the potential. 
We also discuss a possibility of a bound state.
Section \ref{discussion} is devoted to conclusions and discussions. 
Some technical details in the calculations are given in Appendices.
In Appendix \ref{SUGRA}, we derive tree amplitudes of the supergravity for a pair of generally
moving D$p$-branes. 
In Appendix \ref{potential omega-expansion}, we evaluate the one-loop Yang-Mills amplitude
for a system of revolving D$p$-branes, especially $p=3$, by using the $\omega$-expansion. 
The expansion is valid for $\omega<r.$
In Appendix \ref{r/w expansion}, the same amplitude is evaluated by the $r$-expansion, which is valid
for $r<\omega$.

\section{String Threshold Corrections in D-brane Models \label{approximation}}

\vspace{5mm}
We are interested in interaction potential between D-branes, which are relatively moving in a target space-time. 
At weak string coupling,
 we can obtain the potential
 by calculating the one-loop partition function of an open string stretched between the D-branes. 
For simple cases,
 we can quantize the stretched open strings 
 and determine a closed form of the one-loop effective potential. 
But in many other cases where D-branes are accelerating,
 it is not possible to write the effective potential in a closed form,
 since open strings have complicated boundary conditions. 
For example, when two D-branes are revolving like a binary star, 
 open string spectrum can be solved only perturbatively
 with respect to the relative velocity \cite{Iso:2018cwb}. 
Thus, in order to calculate the potential between these D-branes, it is necessary 
to develop an alternative method. 
In this section, 
we propose an efficient method to obtain the interaction potential between generally moving D-branes, 
 including threshold corrections of massive open string modes.
The method was  indicated in a seminal paper \cite{Douglas:1996yp}.

Schematically, the effective potential $V(R)$ is given as 
\begin{equation} 
V(R)\ =\ -\int_0^\infty\frac{dt}{t}\,e^{-\frac{R^2}{2\pi\alpha'}t}Z(t), 
   \label{V(r) schematic}
\end{equation}
 where $R$ is the distance between the D-branes. 
$Z(t)$ is the partition function of the stretched open string with the modulus (Schwinger parameter) $t$, 
 where the factor $e^{-\frac{R^2}{2\pi\alpha'}t}$ due to the string tension is extracted. 
In many known examples
 the $R$-dependence only appears through $e^{-\frac{R^2}{2\pi\alpha'}t}$ and $Z(t)$ is $R$-independent, but generally it is not the case. 
The method for analyzing the effective potential at all ranges of $R$
 is based on a simple idea of 
 separating the integration region into the UV region of $t \in [0,1]$ and the IR region  of $t \in[1, \infty)$.
The IR region for the open strings is dominantly given by the massless modes of open strings. 
If the modular transformation for $Z(t)$ can be explicitly performed, 
 the UV region is mapped to the IR region of the dual closed strings and 
 thus determines the large $R$ behavior of the potential.  
 But as we will see in the next section \ref{sec:stringthreshold}, 
  it may also give sizable contributions to the small $R$ behavior of the potential.  
They are the threshold corrections of infinitely many open string massive modes. 

The UV region $[0,1]$ is dominantly described by the 
 massless closed string modes, i.e., supergravity. 
The property holds even when the modular transformation is not explicitly given.
Thus the open string one-loop amplitude of the UV region is approximated 
by using the supergravity calculations with an appropriate cutoff corresponding to 
 $t \in [0,1]$.

\vspace{5mm}

\subsection{Why are the string threshold corrections important?} \label{sec:stringthreshold}

\vspace{5mm}
In this section, we explain the method of {\it partial modular transformation}.
It can provide a good approximation of the effective potential
without directly performing the one-loop open string amplitude.  
Let us start from  a toy example in the bosonic string theory. 
The model contains an open string tachyon and the potential is not well-defined for small $R$, 
but still it is a good example to see its efficiency and  usefulness of the method. 
The effective potential of a pair of static parallel D$p$-branes in the bosonic string theory 
 is given by
\begin{equation}
V(R)\ =\ -\int_0^\infty\frac{dt}{t}\,
          e^{-\frac{R^2}{2\pi\alpha'}t}(8\pi^2\alpha't)^{-\frac12(p+1)}\eta(it)^{-24}. 
\label{V(r) bosonic}
\end{equation}
The integral contains contribution from the tachyon which makes the integral divergent at small $R$.
We simply ignore it here.
In the following sections,
 we will consider tachyon-free models whose effective potential is well-defined for all ranges of $R$. 

First, let us consider the potential at large $R$. 
As usual,
 the asymptotic behavior of $V(R)$ at large $R$ can be easily determined by using the modular transformation. 
Due to the exponential factor $e^{-\frac{R^2}{2\pi\alpha'}t}$, 
 small $t$ region dominantly contributes to the behavior at large $R$.  
After a modular transformation, we get 
\begin{equation}
V(R)\ =\ -(8\pi^2\alpha')^{-\frac12(p+1)}\int_0^\infty ds\,
         e^{-\frac{R^2}{2\pi\alpha'}s^{-1}}s^{\frac12(p-25)}\eta(is)^{-24}. 
\end{equation}
The large $s$ region gives the dominant contribution at large $R$.
Thus we expand the Dedekind eta function $\eta(is)$ as 
\begin{equation}
\eta(is)^{-24}\ =\ \sum_{n=-1}^\infty d_ne^{-2\pi ns}, 
   \label{Dedekind eta}
\end{equation}
 where $d_{-1}=1$, $d_0=24$, $d_1=324$, $d_2=3200$ etc., and we retain only terms with small $n$. 
Again we ignore the closed tachyon contribution ($n=-1$) here. 
The $n=0$ term gives 
\begin{equation}
V(R)\ \sim\ -(4\pi)^{-\frac12(p+1)}(2\pi\alpha')^{11-p}\Gamma({\textstyle \frac{23-p}2})R^{p-23}, 
\end{equation}
 which is a good approximation for large $R$, up to the tachyonic contribution.
It corresponds to the exchange of massless closed string states, i.e., the dilaton and the graviton. 

The behavior of $V(R)$ at small $R$, however, is more non-trivial. 
Similarly we can expand the $\eta(it)$ in eq.(\ref{V(r) bosonic})
 by using the formula of eq.(\ref{Dedekind eta}). 
Then, discarding the open string tachyon ($d_{-1}$),
 we may think that only the massless open string modes contribute to the behavior of $V(R)$ at small $R$.  
But actually it is not the case because 
 all values of $t$, including large $t$, can contribute to the integral\footnote{
It is known that a singular behavior of physical quantities
 can be extracted solely from the lightest open string states
 which become massless in the singular limit \cite{Douglas:1996yp}. 
}. 
For example,
 the contribution from $n$-th excited states with the coefficient $d_n$ in (\ref{Dedekind eta})
 gives the following contribution to the effective potential $V(R)$;
\begin{eqnarray}
&& \int_{1/\Lambda^2}^\infty\frac{dt}{t}e^{-\frac{R^2}{2\pi\alpha'}t}t^{-\frac12(p+1)}e^{-2\pi nt}
\nonumber \\ [2mm]
&=& \left\{
\begin{array}{lc}
\displaystyle{2\Lambda-\sqrt{4\pi x}} + {\cal O}(1/\Lambda), & (p=0) \\ [2mm] 
\displaystyle{\frac{\Lambda^4}{2}-\Lambda^2 x+\frac{3-2\gamma}{4}x^2-\frac12x^2\log(x/\Lambda^2)}
 + {\cal O}(1/\Lambda^2), & (p=3)
\end{array}
\right.
   \label{bosonic each order}
\end{eqnarray}
where we have introduced the UV cutoff $\Lambda$ (in unit of the string scale) and 
\begin{equation}
x\ :=\ \frac{R^2}{2\pi\alpha'}+2n\pi. 
\end{equation} 
The first and the second terms in the formula for $p=3$
 are nothing but the quartic and quadratic divergences in $d=4$ quantum field theories. 
The third and the fourth terms are the Coleman-Weinberg effective potential 
 with a mass squared, $M^2=x$. 
Since $x$ increases with increasing $n$,
 massive open string modes give huge contributions to the low energy effective potential. 
Therefore, we cannot simply discard the contributions from massive open string states 
 in determining the behavior of $V(R)$ for small $R$, even though they are heavy. 
We also need to take an appropriate treatment of the UV cutoff $\Lambda$ appearing in the above formulas,
 which causes ambiguities of  finite renormalizations of low energy observables.

In addition, the above calculations can provide  behaviors of the potential $V(R)$ only for 
 small $R$ or  large $R$ regions.
But, we are interested in the behavior of potential $V(R)$ in the whole ranges of $R$. 
In the next section,
 we propose an efficient method to evaluate 
  $V(R)$ interpolating the small $R$ and large $R$ regions.

\vspace{5mm}

\subsection{Partial modular transformation}

\vspace{5mm}

We will now provide an efficient method to obtain a good approximation of $V(R)$ for all ranges of $R$. 
Interestingly,
 this method also resolves the issue of the UV divergences mentioned in the previous section. 

Our method is based on the following rewriting of the potential of eq.(\ref{V(r) bosonic}):  
\begin{eqnarray}
V(R) 
&=& -(8\pi^2\alpha')^{-\frac12(p+1)}\left[ \int_1^\infty\frac{dt}{t}\,
    e^{-\frac{R^2}{2\pi\alpha'}t}t^{-\frac12(p+1)}\eta(it)^{-24} \right. 
\nonumber \\ [1mm]
& & \left. \hspace*{2.5cm} +\int_1^\infty ds\,
    e^{-\frac{R^2}{2\pi\alpha'}s^{-1}}s^{\frac12(p-25)}\eta(is)^{-24} \right]. 
\label{V(r) bosonic separate}
\end{eqnarray}
Here, we divided the integration region $[0,\infty)$ for $t$ into $[0,1]$ and $[1,\infty)$,
 and perform the modular transformation for the first half region. 
An advantage of this rewriting is that, since $t,s\ge1$ are satisfied,
 the Dedekind eta functions in the right-hand side
 can be replaced with a few terms in eq.(\ref{Dedekind eta})
 corresponding to light open (closed) string states, even for small $R$. 
For example, 
\begin{equation}
\int_1^\infty\frac{dt}{t}\,
 e^{-\frac{R^2}{2\pi\alpha'}}t^{-\frac12(p+1)}\eta(it)^{-24}\
 \to\ \mbox{(tachyon)}+24\int_1^\infty\frac{dt}{t}\,e^{-\frac{R^2}{2\pi\alpha'}}t^{-\frac12(p+1)}
\label{open approx}
\end{equation}
 is a good approximation for all ranges of $R$. 

Accuracy of the approximation can be estimated as follows. 
Using the expansion of eq.(\ref{Dedekind eta}),
 the left-hand side of eq.(\ref{open approx}) can be estimated as 
\begin{equation}
\sum_{n=0}^\infty d_n\int_1^\infty\frac{dt}{t}\,
    e^{-\frac{R^2}{2\pi\alpha'}t}t^{-\frac12(p+1)}e^{-2\pi nt}\
 <\
\sum_{n=0}^\infty d_ne^{-2\pi n}\int_1^\infty\frac{dt}{t}\,
    e^{-\frac{R^2}{2\pi\alpha'}t}t^{-\frac12(p+1)}. 
\end{equation}
Since $e^{-2\pi}=0.001867$ is a very small number,
 the contributions from massive states are much smaller than those from the massless states. 
One might be worried that the exponential growth of $d_n$ would invalidate this argument. 
However, it is known that $d_n$ grows as $e^{4\pi\sqrt{n}}$,
 which is not large enough to overcome the suppression factor $e^{-2\pi n}$. 
The total contribution (without tachyon) to $V(R)$
 turns out to be smaller than the massless state contribution times an infinite sum 
\begin{equation}
\sum^\infty_{n=0} d_ne^{-2\pi n}\ 
=\ \eta(i)^{-24}-e^{2\pi}\ 
=\ 1.026\,d_0. 
\end{equation}
Therefore, the error due to discarding all massive open string states is less than 3\%. 
Note that the smallness of the error is assured because we have introduced the cutoff at $t=1$. 

The second half in eq.(\ref{V(r) bosonic separate}) can be similarly approximated as 
\begin{equation}
\sum_{n=0}^\infty d_n\int_1^\infty ds\,e^{-\frac{R^2}{2\pi\alpha'}s^{-1}}s^{\frac12(p-25)}e^{-2\pi ns}\
 <\ 
\sum_{n=0}^\infty d_ne^{-2\pi n}\int_1^\infty ds\,e^{-\frac{R^2}{2\pi\alpha'}s^{-1}}s^{\frac12(p-25)}. 
\end{equation}
Therefore, retaining the contributions from massless closed string states
 gives a good approximation with the same accuracy as above. 
We emphasize that the accuracy of the approximation does not depend on $R$,
 so this approximation is valid for all range of $R$. 
If one needs a more precise approximation,
 one can retain the first excited states for both open and closed string channels. 
Then, 
\begin{equation}
\sum^\infty_{n=1} d_ne^{-2\pi n}\ =\ \eta(i)^{-24}-e^{2\pi}-24\ =\ 1.019\,d_1e^{-2\pi}, 
\end{equation}
shows that the expected error is about $0.019\,d_1e^{-2\pi}/d_0=0.05$\%. 

\vspace{5mm}

Several comments are in order. 
First the method is to sum the contributions from the open massless modes and the closed massless modes.
If we did not introduce the Schwinger parameter cutoff, it would be a double counting. 
But as is clear from the procedure, it is not. 
Next, the expression is finite, as long as 
 the square of the mass of  the ``tachyonic'' state is positive. 
This implies that 
 the issue of the UV divergences and ambiguities of  finite renormalizations
 mentioned above are resolved by summing all open string massive contributions. 
Finally, in eq.(\ref{V(r) bosonic separate}),
 we separated the region of the moduli integration at $t=s=1$,
 which is the fixed point of the modular transformation. 
If we separate the modulus at a different value, $t=2$ and $s=1/2$ for example, 
 the suppression factor for the open string channel becomes $e^{-4\pi}=3.487\times10^{-6}$
 and the approximation becomes better. 
However, the suppression factor for the closed string channel becomes $e^{-\pi}=0.04321$,
 giving a worse approximation. 
Hence the choice $t=s=1$ seems  to be optimal. 

\vspace{5mm}

\subsection{Another example: D3-branes at angle}

\vspace{5mm}

As another example in the superstring case,
we consider a pair of D3-branes at angle in Type IIB string theory. 
We follow the notations of the section 13.4 in \cite{Polchinski:1998rq,Polchinski:1998rr}. 
For $\phi_4=0$, the one-loop effective potential is given by
\begin{equation}
V(R)\ =\ -\int_0^\infty\frac{dt}{t}(8\pi^2\alpha't)^{-\frac12}e^{-\frac{R^2}{2\pi\alpha'}t}
 \frac{i\prod_{a=1}^4\vartheta_{11}(\frac i\pi\phi_a't,it)}
 {\eta(it)^3\prod_{a=1}^3\vartheta_{11}(\frac i\pi\phi_at,it)}, 
\label{D3-at-agnle-string}
\end{equation}
 where 
\begin{eqnarray}
\phi_1'\ :=\ \frac12(\phi_1+\phi_2+\phi_3), & & \phi_2'\ :=\ \frac12(\phi_1+\phi_2-\phi_3), \nonumber \\
\phi_3'\ :=\ \frac12(\phi_1-\phi_2+\phi_3), & & \phi_4'\ :=\ \frac12(\phi_1-\phi_2-\phi_3). 
\end{eqnarray}
We assume that the angles $\phi_a$ are small
 so that the mass spectrum of the stretched open string
 is not largely deviated from that for the BPS configuration with $\phi_a=0$. 
This integral is convergent for large $t$ if 
\begin{equation}
\sum_{a=1}^4|\phi_a'|\ \le\ \sum_{a=1}^3|\phi_a|
\end{equation}
 is satisfied. 
This corresponds to the condition for the absence of open string tachyons. 
A solution of this condition is 
\begin{equation}
\phi_1\ =\ \phi_2\ =\ \phi_3\ =\ \phi
   \label{no open tachyon}
\end{equation}
 for any $\phi$. 
The integral is always convergent for small $t$ since there is no closed string tachyons. 
Therefore, the effective potential $V(R)$ with $\phi_a$ satisfying eq.(\ref{no open tachyon})
 is well-defined for all ranges of $R$. 

\begin{figure}[!htb]
\center
\includegraphics [scale=.3] {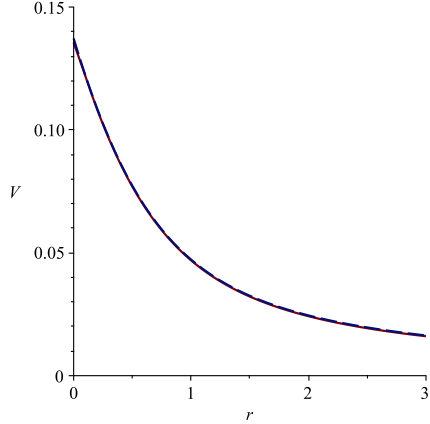}
\vspace{10mm}
\caption{\small
The effective potential $V(R)$ for D3-branes at angle with $\phi=\pi/12$ and $\alpha'=1$.
The exact effective potential in eq.(\ref{D3-at-agnle-string}) is drawn with red solid line.
The blue broken line shows the potential using the approximate formula in eq.(\ref{D3approx}),
 which agrees very well with the exact one.
}
\label{VD3}
\end{figure}
\vspace{5mm}

Similarly to the bosonic example in the previous section,
 the stringy result of eq.(\ref{D3-at-agnle-string})
 can be approximated by a sum of open light and closed massless contributions,
\begin{eqnarray}
\tilde{V}(R) 
&=& \frac1{\sqrt{8\pi^2\alpha'}}\left[ \int_1^\infty dt\,t^{-\frac32}e^{-\frac{R^2}{2\pi\alpha'}t}
\frac{2\sinh(\frac32\phi t)\sinh^3(\frac12\phi t)}{\sinh^3(\phi t)} \right.
\nonumber \\
& & \left. \hspace*{1.5cm} +\int_1^\infty ds\,s^{-\frac32}e^{-\frac{R^2}{2\pi\alpha'}s^{-1}}
\frac{2\sin(\frac32\phi)\sin^3(\frac12\phi)}{\sin^3\phi} \right]. 
\label{D3approx}
\end{eqnarray}
This formula provides a good approximation of eq.(\ref{D3-at-agnle-string}).
Indeed, the plot of $V(R)$ and $\tilde{V}(R)$ is shown in Figure \ref{VD3}.
The error for the approximation is quite small for all range of $R$, which
is difficult to see by naked eyes.

\subsection{General Recipe \label{recipe} }

\vspace{5mm}

Let us summarize the method to give an efficient approximation to 
the one-loop effective potential $V(R)$ at all ranges of $R$.  
In the example in the previous section, 
given a modulus integral for the effective potential of interest in eq.(\ref{D3-at-agnle-string}),
 we divided the integration region into two,
 and performed the modular transformation for one of the integrals. 
Then, we retained only the contributions from light states to the integrals,
 open (nearly) massless states and closed massless states.  
The resulting expression gives a good approximation to the full effective potential for all ranges of $R$. 
Now we generalize the method to more complicated situations. 
To determine the approximate expression for the effective potential in the D-brane system,
 we did not actually need to know the full spectrum of the stretched open string.
Only the information of the effective theories of the open massless states
 and the closed massless states are necessary. 
Namely the approximate effective potential is given as a sum of
the SYM and the supergravity contributions;
\begin{equation}
\tilde{V}(R)\ =\ \tilde{V}_o(R)+\tilde{V}_c(R), 
\end{equation}
 where $\tilde{V}_o(R)$ and $\tilde{V}_c(R)$ are schematically given as 
\begin{eqnarray}
\tilde{V}_o(R) 
&=& -\int_1^\infty\frac{dt}{t}\int\frac{d^{D}k}{(2\pi)^{D}}
     \sum_{\rm light\ open}e^{-2\pi tE_o(k)-\frac{R^2}{2\pi\alpha'}t}, \nonumber \\
\tilde{V}_c(R) 
&=& -\int_1^\infty ds\,
    \sum_{\stackrel{\rm \scriptstyle massless}{\rm closed}}\int\frac{d^{D'}k}{(2\pi)^{D'}}\,
    \langle B|c\rangle\langle c|B'\rangle e^{-2\pi sE_c(k)-\frac{R^2}{2\pi\alpha'}s^{-1}}.
\label{V_approx}
\end{eqnarray}
Here $|B\rangle$ and $|B'\rangle$ are the boundary states for the D-branes and
 $|c\rangle$ are the closed string massless states propagating between D-branes. 

$\tilde{V}_o(R)$ is the Schwinger parametrization of the one-loop determinant
 for light open string states with the UV cutoff at the string scale. 
Thus, it can be obtained from the worldvolume theory of the D-brane system under consideration. 
Suppose that a D-brane configuration of interest
 is described by a classical field configuration in the worldvolume theory. 
Then, the one-loop calculation around the classical configuration gives the desired one-loop determinant. 
If we rewrite this in terms of the Schwinger parameter and put a suitable cutoff,
 we obtain $\tilde{V}_o(R)$ without performing any stringy calculations. 

On the other hand,
 $\tilde{V}_c(R)$ is obtained from the massless closed string exchange between the D-branes.
 For general configurations of D-branes, see \cite{Kazama:1997bc,Hirano:1996pf}. 
This can be understood by noticing that
 the Schwinger parametrization of the massless propagator in $D'$ dimensions is proportional to 
\begin{equation}
\int\frac{d^{D'}k}{(2\pi)^{D'}}\frac{e^{ikx}}{k^2}\
 =\ (4\pi)^{-\frac{D'}2}\int_0^\infty ds\,s^{-\frac{D'}2}e^{-\frac14x^2s^{-1}}. 
\label{Schwinger}
\end{equation}
The interaction vertex $\langle B|c\rangle$ of a D-brane to a closed string state
 is given by the corresponding Dirac-Born-Infeld (DBI) action with Chern-Simons (CS) term,
 provided that the trajectory of the D-brane is specified. 
Then, we obtain $\tilde{V}_c(R)$ by determining the appropriate tree amplitudes in supergravity,
 written in the Schwinger parametrization, and putting a suitable UV cutoff at the string scale. 

\vspace{5mm}

Now, we have the recipe for a well-approximated expression 
 to the full one-loop effective potential of a D-brane system, which
 includes threshold contributions from infinitely many massive open string modes: 
\begin{enumerate}
\item 
Find a classical configuration in the worldvolume theory of a D-brane system under consideration, 
 which corresponds to the D-brane configurations we are interested in. 
Then perform one-loop calculations around the classical configuration, 
 and express the resulting one-loop determinant in terms of the Schwinger parameter $t$. 
UV cutoff in the $t$-integration is introduced. 
\item 
Calculate the classical potential, mediated by the massless closed string states in supergravity,
 between the given configurations of D-branes. 
The coupling vertices are derived from the corresponding DBI action with CS term. 
Express the result in terms of the Schwinger parameter $s$ and introduce the UV cutoff. 
\item 
Normalize $t$ and $s$ such that the $R$-dependence appears in either of the form 
\begin{equation}
e^{-\frac{R^2}{2\pi\alpha'}t}, \hspace{1cm} e^{-\frac{R^2}{2\pi\alpha'}s^{-1}}, 
\end{equation}
 and put the ``cutoff'' at $t,s=1$. This corresponds to introducing the UV cutoff 
 at the string mass scale $m_s=(2 \pi \alpha')^{-1/2}$.
\item 
The sum of the above two expressions
 gives a good approximation $\tilde{V}(R)$ to the full effective potential $V(R)$ for all ranges of $R$. 
\end{enumerate}

\vspace{5mm}

If we interpret this recipe from the open string channel,
 what we have done amounts to summing all the stringy threshold corrections
 to the effective potential with a very good accuracy. 
This can be done by converting the threshold corrections
 into a contribution from the closed string massless states. 
The open-closed duality plays a key role in this calculation. 


Note that the above calculations in the recipe can be performed even for {\it off-shell} configurations of D-brane system. 
In principle, an off-shell interaction of D-branes would be calculated in terms of an open string {\it field} theory. 
The leading order contribution with respect to $g_s$ is given by the one-loop open string amplitude. 
The Schwinger parametrization for this amplitude should be available. 
According to the recipe given above, we divide the integration region of the Schwinger parameter into two. 
A half of them corresponding to small Schwinger parameters can be replaced with a tree amplitude of a closed string {\it field} theory, assuming that the open-closed duality persists off-shell. 
By truncating both string field theory calculations, they are reduced to off-shell calculations in a worldvolume theory of the D-branes and in a supergravity. 
We know, in principle, how to perform the both calculations. 
These calculations should give an approximate effective potential for an off-shell D-brane system. 

We should also emphasize again that
 our recipe does not suffer from a double-counting problem,
 since we have introduced a cutoff in the Schwinger parameter integration. 
One can convince oneself of the validity of our recipe
 by examining the large $R$ behavior of $\tilde{V}_o(R)$. 
Due to the factor $e^{-\frac{R^2}{2\pi\alpha'}t}$ and the cutoff at $t=1$, 
 $\tilde{V}_o(R)$ exponentially damps at large $R$, $V(R) \sim e^{-R^2/2\pi\alpha'}$.
Therefore, the Newton potential appears only from $\tilde{V}_c(R)$. 
It is also important to notice that, although  $\tilde{V}_c(R) \sim -1/R^{7-p}$  at large $R$, 
 $\tilde{V}_c(R)$ is finite in the limit $R\to0$, due to the cutoff at $s=1$. 

\vspace{5mm}

In the following, we apply our  method to the revolving D-branes. 
In \cite{Iso:2018cwb}, we investigated this system based on a worldsheet theory of the stretched open string. 
However, there were several difficulties. 
One of them is the fact that
 the revolving configuration is off-shell at tree level,
 so that worldsheet calculations could have some troubles,
 for example an ambiguity for the renormalization procedure. 
We will see that
 our method in this paper gives a quite reasonable finite result for the effective potential
 of the revolving D-branes, improving our previous investigation in \cite{Iso:2018cwb}.



\vspace{1cm}

\section{Gauge theory calculations in revolving D$p$-branes }
\label{revolving Dp}
\vspace{5mm}

In the following sections, 
we apply the recipe in the previous section to 
calculate the effective potential $\tilde{V}(R)$
 for a system of D$p$-branes revolving around each other.
The distance $R=2r$ is chosen to have mass dimension 1 and so is the radius of the revolution $r$.
The radius with dimension -1 is given by $2 \pi \alpha' r$.
In this section, we set $2\pi\alpha'=1$.
To determine $\tilde{V}_o(2r)$,
 we perform a one-loop calculation around a suitable background field configuration
 in maximally supersymmetric Yang-Mills theory in $p$+1 dimensions\footnote{
The effective theory of D$p$-branes
 is given by the DBI action with CS term and contains higher 
 derivative corrections to the SYM theories.
These higher dimensional vertices are suppressed by a factor {$1/m_s$}
 and the corrections to the effective  potential between D-branes can be neglected in the region
 {$r < m_s$}. 
Such suppression property is different from the threshold corrections of massive open string modes 
 running in internal lines of the Feynman diagrams,
 which may give sizable contributions to the potential as discussed in section \ref{sec:stringthreshold}.
 }. 
For $\tilde{V}_c(2r)$, in the next section,
 we calculate the amplitude for the one-particle exchange
 between the revolving D$p$-branes in Type II supergravity.

Before discussing details, we first briefly explain our procedure to calculate the one-loop radiative
corrections to the effective potential based on the background field method.
We first divide the field configurations into collective coordinates and 
fluctuations around it. The collective coordinates represent generally off-shell background.
In the present case, it represents the revolving motion of D-branes.
On the other hand, fluctuations are chosen so that they are perpendicular to the collective coordinates in the field configurations
so that the Gaussian integrations can be performed. Massive open string modes correspond to these
fluctuations. 
Then by calculating the one-loop determinant, we can obtain one-loop corrections to the effective potential.

The maximally supersymmetric Yang-Mills theory in $p$+1 dimensions
consists of gauge fields $A_{\mu}$ $(\mu=0,1,\cdots,p)$,
 scalar fields $\Phi_I$ $(I=p+1,\cdots,9)$ and fermions
 which are obtained from a Majorana-Weyl fermion $\Psi$ in ten dimensions via the dimensional reduction. 
Since we want to describe two D$p$-branes revolving around each other, 
 we choose the gauge group to be ${\rm SU}(2)$.
The ${\rm U}(1)$ part describes the center of mass degrees of freedom
 and it is irrelevant in the present analysis.
The signature of the metric is $(-1,1,\cdots,1)$ all through the paper.

The action including a gauge-fixing term and the associated ghost action is given by
\begin{eqnarray}
S 
&=&\frac{1}{g^2}\int d^{p+1}x\
 \textrm{Tr}\left[-\frac14F_{\mu \nu}F^{\mu \nu}-\frac12D_{\mu}\Phi_ID^{\mu}\Phi^I
 +\frac14([\Phi_I,\Phi_J])^2\right.\nonumber \\
&&\left.\hspace{2.5cm}+\frac i2\bar \Psi \Gamma^{\mu}D_{\mu}\Psi
 +\frac12\bar \Psi \Gamma^I[\Phi_I,\Psi]\right] \nonumber \\
& &-\frac1{2g^2}\int d^{p+1}x\ \textrm{Tr}\left[(\partial^\mu A_\mu-i[B^I,\Phi_I])^2\right] \nonumber \\ [1mm]
& &+\frac1{g^2}\int d^{p+1}x\ \textrm{Tr}\left[\bar c(\partial^\mu D_\mu c-[B^I,[\Phi_I,c]])\right], 
\label{SYM action}
\end{eqnarray}
 where $\Gamma^\mu, \Gamma^I$ are the Dirac matrices in ten dimensions,
  and $B_I$ are background fields for the scalars $\Phi_I$. 
We have chosen the background field gauge 
\begin{equation}
\partial^\mu A_\mu-i[B_I,\Phi^I]=0.
\end{equation}
(See e.g. \cite{Becker:1997wh}.)
This is a natural gauge choice
 from the point of view of ${\cal N}=1$ SYM theory in ten dimensions.

\vspace{5mm}
\subsection{SYM in a general background $B_I$}

\vspace{5mm}

We expand the action of eq.(\ref{SYM action}) around the background $B_I$ by setting
\begin{equation}
A_\mu=a_\mu,\hspace{1cm}\Phi_I=B_I+\phi_I,\hspace{1cm}\Psi=\psi.
\end{equation} 
The relevant part $S_2$ of the action for obtaining the one-loop determinant
 consists of terms quadratic in the fluctuations $a_\mu$, $\phi_I$ and $\psi$. 
It is given by 
\begin{eqnarray}
S_2&=&\frac{1}{g^2}\int d^{p+1}x\
 \textrm{Tr}\left[-\frac12(\partial_\mu a_\nu)^2+\frac12[B_I,a_\mu]^2-\frac12(\partial_\mu \phi_I)^2
 +\frac12[B_I,\phi_J]^2 \right.
\nonumber \\ [2mm]
&&\left. \hspace{3cm}+[B_I,B_J][\phi^I,\phi^J]+2i\partial_\mu B_I[a^\mu,\phi^I]\right.
\nonumber \\ [2mm]
&&\left.\hspace{2.5cm}+\frac i2\bar \psi \Gamma^{\mu}\partial_{\mu}\psi+\frac12\bar \psi \Gamma^I[B_I,\psi]
+\bar c \partial^\mu \partial_\mu c-\bar c[B^I,[B_I,c]]\right].
\nonumber \\
\end{eqnarray}

In the following,
 we are interested in a background configuration $B_I$
 corresponding to a motion of the D$p$-branes. 
The background configuration $B_I$ takes the form of
\begin{equation}
B_I=b_I(t) \sigma_3,
\end{equation}
 where $b_I(t)$ are functions of time $t$,
 describing the trajectories of the D$p$-branes, and $\sigma_i$ are Pauli matrices. 
It turns out that most of the terms in $S_2$ including $B_I$
 are the mass terms for the fluctuations.
In addition,
 there is a mixing term of the gauge field $a^\mu$ and the scalar field $\phi^I$,
 $2i\partial_\mu B_I[a^\mu, \phi^I]$ which gives a non-trivial effect of the background
 to the one-loop determinant. 

The effective potential between the D$p$-branes are induced by an open string stretched between them. 
Such an open string corresponds to the off-diagonal components of the fluctuations,
 which are proportional to $\sigma_{1,2}$. 
As mentioned before, there are no linear terms for them in the action. 
To compute the one-loop determinant relevant for the effective potential,
 we perform the Wick rotation
\begin{equation}
 t=-i\tau, \hspace{1cm} a_0=ia_\tau, \hspace{1cm} \Gamma^0=-i\Gamma_\tau
\label{Wick rotation}
\end{equation}
to regularize the path integral, and set 
\begin{equation}
a_m =\tilde{a}_m\sigma_+ + \tilde{a}_m^\dagger \sigma_-, 
\hspace{1cm}
 \phi_I=\varphi_I \sigma_+ + \varphi_I ^\dagger \sigma_-,
\hspace{1cm}
 \psi=\chi\sigma_+ + \tilde {\chi}\sigma_-,
\end{equation}
 where $\sigma_\pm:=\frac12(\sigma_1\pm i\sigma_2)$ and $m=1,\cdots,p,\tau$. 
Note that $\chi$ and $\tilde{\chi}$ are related to each other by the Majorana-Weyl condition of $\psi$. 
Inserting them into $S_2$, we obtain 
\begin{eqnarray}
S_2 
&=& \frac1{g^2}\int d^{p+1}x\ {\rm Tr}\left[
 |\partial_m a_n|^2+4(b_I)^2|a_m|^2+|\partial_m\varphi_I|^2+4(b_I)^2|\varphi_J|^2 \right. \nonumber \\
& & \hspace*{2cm}
-4i\partial_m b_I(a_m\varphi_I^\dagger-a_m^\dagger \varphi_I)
+i\bar \chi \Gamma_m \partial_m\chi - 2\bar \chi\Gamma_Ib_I\chi
\nonumber \\ [1mm]
& & \left. \hspace*{2cm}+\bar c_+\partial^2 c_+ + \bar c_-\partial^2 c_--4(b_I)^2(\bar c_+ c_+ + \bar c_- c_-)
 \right], 
\end{eqnarray}
where we denoted $a_m$ instead of $\tilde{a}_m$ for notational simplicity. 

\vspace{5mm}

\subsection{One-loop amplitude of SYM in revolving D$p$-branes}

\vspace{5mm}

We now consider a specific background corresponding
 to the revolving D$p$-branes ($p \leq 7$) in the 8-9 plane. 
The extended directions of D$p$-branes are taken to be the same and thus always in parallel. 
The corresponding background configuration is given by 
\begin{equation}
b_8=r \cos \omega \tau,\hspace{1cm}b_9= r \sin \omega \tau, \label{B_I}
\end{equation}
 and $b_I=0$ otherwise,
 where $\omega$ is the angular frequency of the revolution
 and $r$ is the radius of the circle on which the D-branes are revolving. 
Note that $\omega$ above has been analytically continued
 according to the Wick rotation of eq.(\ref{Wick rotation}). 
To recover the results in the Lorentzian signature, we will replace $\omega$ with $-i\omega$. 

At first sight,
 since the quadratic action $S_2$ for the above stationary configuration $b_I$ is $\tau$-dependent,
 one may think that the one-loop determinant also depends on $\tau$.
Indeed, it is the case when D-branes are moving
 with a constant relative velocity \cite{Bachas:1995kx,Lifschytz:1996iq}.
But in the present situation, since the motion is stationary, 
 the $\tau$-dependence of the effective potential can be eliminated. 
By introducing new fields $\varphi_\pm$ defined by
\begin{equation}
\varphi_{\pm}:=\frac1{\sqrt2}e^{\mp i\omega \tau}(\varphi_8\pm \varphi_9), 
\end{equation}
 the $\tau$-dependence of the bosonic part of $S_2$ can be eliminated. 
Similarly, the $\tau$-dependence of the fermionic part of $S_2$ can be eliminated by introducing 
\begin{equation}
\theta:=\exp \left[\frac12 \omega \tau \Gamma^{89}\right]\chi.
\end{equation}
In terms of these new fields, the quadratic action $S_2$ becomes 
\begin{equation}
S_2=\frac1{g^2}\int d^{p+1}x\,\left[ L_B+L_F+L_{\rm free} \right]
\end{equation}
 where 
\begin{eqnarray}
L_B
&=&|(\partial_m+i\omega_m)\varphi_+|^2+4r^2|\varphi_+|^2+|(\partial_m-i\omega_m)\varphi_-|^2+4r^2|\varphi_-|^2
\nonumber \\
&&+|\partial_m a|^2+4r^2|a|^2
 -2\sqrt{2}r\omega
  \left(\varphi_- a^\dagger+\varphi_-^\dagger a-\varphi_+ a^\dagger -\varphi_+^\dagger a \right),
\\
L_F
&=&i\bar{\theta}\Gamma_m \left( \partial_m-\frac12\omega_m\Gamma^{89} \right)\theta
 -2r\bar{\theta}\Gamma^8\theta, \\
L_{\rm free}
&=&|\partial_m a_i|^2+4r^2|a_i|^2
 +\bar{c}_+\partial^2c_+-4r^2\bar{c}_+c_++\bar{c}_-\partial^2c_--4r^2\bar{c}_-c_-.
\nonumber \\
\end{eqnarray}
Here, we defined $\omega_m:=\omega\delta_{m\tau}$ and $i=1,2,\cdots,p$. 
For notational simplicity, we used $a$ instead of $a_\tau$. 

Now we can compute the one-loop determinant. 
Since the $\tau$-dependence is no longer present, we can employ the momentum representation. 
Then, the bosonic Lagrangian $L_B$ can be written as 
\begin{eqnarray}
&&(k^2+4r^2)\left|a(k)-\frac{2\sqrt2r\omega}{k^2+4r^2}(\varphi_-(k)-\varphi_+(k))\right|^2
\nonumber \\ [2mm]
&&+(k^2+\omega^2+4r^2+2\omega k_\tau)|\varphi_+(k)|^2+(k^2+\omega^2+4r^2-2\omega k_\tau)|\varphi_+(k)|^2
\nonumber \\ [2mm]
&&-\frac{8(r\omega)^2}{k^2+4r^2}|\varphi_-(k)-\varphi_+(k)|^2, 
\end{eqnarray}
where $k^2=(k_m)^2$. 
The path integral for $a$ can be easily performed,
 resulting in the determinant $\det(-\partial^2+4r^2)^{-1}$. 
To perform the path integral for $\varphi_\pm$, we need to diagonalize the matrix 
\begin{equation}
\left(k^2+\omega^2+4r^2-\frac{8(r\omega)^2}{k^2+4r^2}\right) I_{2 \times 2}+
\left(\begin{array}{cc} \displaystyle{2\omega k_\tau}&\displaystyle{\frac{8(r\omega)^2}{k^2+4r^2}} \\ 
\displaystyle{\frac{8(r\omega)^2}{k^2+4r^2}} &\displaystyle{-2\omega k_\tau} \end{array}\right),
\end{equation}
where $I_{2\times 2}$ is the diagonal matrix. 
Its eigenvalues are given by
\begin{equation}
E_{B\pm}(k):=k^2+\omega^2+4r^2-\frac{8(r\omega)^2}{k^2+4r^2}
 \pm\sqrt{4\omega^2k_\tau^2+\left(\frac{8(r\omega)^2}{k^2+4r^2}\right)^2}.
\end{equation}
Hence, the path integral for the bosonic field $\varphi_\pm$
gives $\det(E_{B+}(-i\partial))^{-1}\det(E_{B-}(-i\partial))^{-1}$. 

\vspace{5mm}

Next, consider the fermionic part $L_F$. 
In the momentum representation, it can be written as 
\begin{eqnarray}
-(\begin{array}{cc} \bar \theta_+&\bar \theta_- \end{array})
\left(\begin{array}{cc} \Gamma_m(k_m+\frac12 \omega_m)&2r \Gamma^8\\
2r\Gamma^8&\Gamma_m(k_m-\frac12 \omega_m) \end{array}\right)
\left(\begin{array}{c} \theta_+\\ \theta_- \end{array}\right), 
\label{quadratic fermion}
\end{eqnarray}
 where $\theta_\pm$ satisfy $i\Gamma^{89}\theta_\pm=\pm\theta_{\pm}$. 
The result of the path integral is given by the determinant of the following matrix 
\begin{eqnarray}
& & \left[
\begin{array}{cc}
\left( k_m+\frac12\omega_m \right)^2+4r^2 & 2r\omega\Gamma^{\tau8} \\
2r\omega\Gamma^{8\tau} & \left( k_m-\frac12\omega_m \right)^2+4r^2
\end{array}
\right] \nonumber \\
&=& \left[
\begin{array}{cc}
1 & 0 \\
0 & \Gamma^{8\tau}
\end{array}
\right]\left[
\begin{array}{cc}
\left( k_m+\frac12\omega_m \right)^2+4r^2 & 2r\omega \\
2r\omega & \left( k_m-\frac12\omega_m \right)^2+4r^2
\end{array}
\right]\left[
\begin{array}{cc}
1 & 0 \\
0 & \Gamma^{\tau8}
\end{array}
\right], \nonumber \\
\end{eqnarray}
 which is the square of the matrix in eq.(\ref{quadratic fermion}). 
The eigenvalues of this matrix are 
\begin{equation}
E_{F\pm}(k)=k^2+\frac14 \omega^2+4r^2\pm \omega \sqrt{k_\tau^2+4r^2}
\end{equation}
with multiplicity four for each of them. 
Therefore, the resulting determinant is given by
 $\det(E_{F+}(-i\partial))^4\det(E_{F-}(-i\partial))^4$. 

\vspace{5mm} 

The remaining part $L_{\rm free}$ simply gives $\det(-\partial^2+4r^2)^{-5}$. 

\vspace{5mm} 

In summary, we obtain the one-loop determinant  whose logarithm is given by
\begin{eqnarray}
&&\log \Bigl[ \det(-\partial^2+r^2)^{-6}\det(E_{B+}(-i\partial))^{-1}\det(E_{B-}(-i\partial))^{-1} \nonumber \\
&& \hspace*{1cm}\times \det(E_{F+}(-i\partial))^{4}\det(E_{F-}(-i\partial))^{4} \Bigr] \nonumber \\ [2mm]
&=& 
\int_{\Lambda^{-2}}^\infty \frac{dt}{t}\int \frac{d^{p+1}k}{(2\pi)^{p+1}}
\Bigl[ e^{-tE_{B+}(k)}+e^{-tE_{B-}(k)} \nonumber \\ [1mm]
&&\hspace*{2cm}-4\left( e^{-tE_{F+}(k)}+e^{-tE_{F-}(k)} \right)+6\,e^{-t(k^2+4r^2)} \Bigr], 
   \label{1-loop det}
\end{eqnarray}
where $\Lambda$ is a UV momentum cutoff with mass dimension 1.
In section \ref{VO}, it is fixed at $\Lambda=m_s$
following the recipe in section \ref{recipe}.
 
A similar calculation was performed in \cite{Bachas:1995kx} where D0-branes are moving with constant velocities. 
In this situation, an open string stretched between the D0-branes changes its length with time. 
If the change is non-adiabatic, 
this causes the parametric resonance, resulting in open string pair productions. 
Indeed, the one-loop determinant for this system has an imaginary part. 
On the other hand, since our investigation is performed to 
find a possibility of a solution in which the induced attractive
potential and the centrifugal potential are balanced, 
we assumed that 
the revolving D-brane system we have discussed so far is stationary. 
Thus the length of the stretched open string is constant in time and
there is no pair production of open strings, indicated by the absence of an imaginary part in the one-loop determinant (\ref{1-loop det}). 

\vspace{5mm}

\subsection{One-loop effective potential $\tilde{V}_o(2r)$ from SYM \label{VO} }

\vspace{5mm}

The contributions from the open light modes to the effective potential $\tilde{V}_o(2r)$
 are given as a sum of the bosonic and fermionic ones,
\begin{equation}
\tilde{V}_o(2r)=\tilde{V}_{o,B}(2r)+\tilde{V}_{o,F}(2r)
\end{equation}
 where they are given by
\begin{eqnarray}
\tilde{V}_{o,B}(2r)
&=&-\int_{\Lambda^{-2}}^\infty\frac{dt}{t}\int\frac{d^{p+1}k}{(2\pi)^{p+1}}
\left[ e^{-tE_{B+}(k)}+e^{-tE_{B-}(k)}+6e^{-t\left( k^2+4r^2 \right)} \right],
\nonumber \\ \\
\tilde{V}_{o,F}(2r)
&=& 4\int_{\Lambda^{-2}}^\infty\frac{dt}{t}\int\frac{d^{p+1}k}{(2\pi)^{p+1}}
\left[ e^{-tE_{F+}(k)}+e^{-tE_{F-}(k)} \right]. 
\end{eqnarray}
The ghost contribution is included in the bosonic part, $\tilde{V}_{o,B}(2r)$. 

Let us now determine the cutoff parameter  $\Lambda$ following the recipe in section \ref{recipe}. 
The factor due to the string tension in the above expression is 
 ${\rm exp}(-r^2t/(\pi\alpha')^2)$ where $\alpha'$ is recovered.
 Since $R=2r$, 
 the recipe tells us to choose the cutoff at $\tilde{t}_{\rm cutoff}=1$ when we rescale the variable
 $t$ so that ${\rm exp}(-r^2t/(\pi\alpha')^2) = {\rm exp}(-(2r)^2\tilde{t}/2 \pi\alpha')$.
Thus we choose $t=2 \pi \alpha' \tilde{t}$ and 
 the momentum cutoff $\Lambda$ can be fixed 
 by the relation, $t_{\rm cutoff} =\Lambda^{-2} = 2\pi\alpha' \tilde{t}_{\rm cutoff} = 2\pi\alpha'$.
Therefore, $\Lambda=m_s$.

Though $\Lambda$ should be fixed as above, 
 it is interesting to see the asymptotic behavior of $\tilde{V}_o(2r)$
 at large $r$ in the limit $\Lambda\to\infty$. 
By rescaling the integration variables, $\tilde{V}_o(2r)$ is rewritten as
\begin{eqnarray}
\tilde{V}_o(2r)
&=&-r^{p+1}\int_{r^2\Lambda^{-2}}^\infty\frac{dt}{t}\int\frac{d^{p+1}k}{(2\pi)^{p+1}}e^{-t(k^2+4)} 
\Bigg[ 6 - 8e^{-\frac{\alpha^2}4t}\cosh\left( t\alpha\sqrt{k_\tau^2+4} \right) 
\nonumber \\
&& \hspace*{1cm}+ 
 2e^{-t\left( \alpha^2-\frac{8\alpha^2}{k^2+4} \right)}
 \cosh\left( t\sqrt{4\alpha^2k_\tau^2+\left( \frac{8\alpha^2}{k^2+4} \right)^2} \right)  \Bigg], 
\end{eqnarray}
 where $\alpha:=\omega/r$. 
This indicates that the $1/r$ expansion of this expression corresponds to the $\alpha$ expansion. 
We find that there are no terms with an odd power of $\alpha$, 
 as it should be,
 since the potential is independent of the direction of rotation with angular frequency $\omega$.
The ${\cal O}(\alpha^0)$ terms cancel trivially due to supersymmetry. 
The next ${\cal O}(\alpha^2)$ terms also cancel between $\tilde{V}_{o,B}(2r)$ and $ \tilde{V}_{o,F}(2r)$;
\begin{equation}
-r^{p+1}\int_{r^2\Lambda^{-2}}^\infty\frac{dt}{t}\int\frac{d^{p+1}k}{(2\pi)^{p+1}}e^{-t(k^2+4)}
 \cdot 16\alpha^2\left( \frac t{k^2+4}-t^2 \right)\stackrel{\Lambda\to0}{\longrightarrow}0. 
\end{equation}
Then, the leading non-vanishing terms
 are  ${\cal O}(r^{p+1}\alpha^4)$, 
 or equivalently ${\cal O}(v^4/r^{7-p})$, where $v:=r\omega$. 
This behavior,
 which can be interpreted as the effective potential for D$p$-branes at large $r$,
 is the same as the one expected from the supergravity calculation,
 which will be shown in the next section.

The effective potential $\tilde{V}_o(2r)$ in the Lorentzian signature
 is obtained by the replacement $\omega \rightarrow -i \omega$ after evaluating the integral
 in the Euclidean signature. 
Details are discussed in section \ref{Epotential}. 
We briefly comment on some properties of the effective potential. 
For $r > l_s$, the effective potential 
represents an attractive force, 
which qualitatively agrees with the supergravity result.
For small $r < l_s$, on the other hand, 
the effective potential behaves nontrivially as a function of $r$ and $\omega$. 
Many cancellations occur between bosons and fermions and
we will show  that, for $p=3$, a minimum of the potential appears 
at a fixed value of $\omega$.

\vspace{5mm}

\section{Supergravity calculations in revolving D$p$-branes} 
\label{sugra}

\vspace{5mm}

In this section,
 we calculate the classical potential $\tilde{V}_c(2r)$ 
 by the one-particle exchanges of massless closed string modes. 

\vspace{5mm}

\subsection{Potential between D-branes mediated by supergravity fields} \label{propagator}
 \vspace{5mm}
 
The relevant fields are the graviton, dilaton and  R-R ($p$+1)-field. 
The bosonic part of the action of Type II supergravity is given by
\begin{equation}
S_{\rm SUGRA}=\frac1{2\kappa_{10}^2}\int d^{10}x
\sqrt{-g}\left[R+\frac12(d \Phi)^2+\frac12(dC^{(p+1)})^2+\cdots \right], 
\end{equation}
 where the fields are normalized such that the kinetic terms become canonical. 
Then the propagators are given by 
\begin{eqnarray}
\mbox{dilaton:} && \Delta(x):=2\kappa_{10}^2\int \frac{d^{10}k}{(2\pi)^{10}}\frac{e^{ik\cdot x}}{k^2}, \\
\mbox{graviton:} && \Delta_{\mu\nu;\rho\sigma}(x):=\left(\eta_{\mu\rho}\eta_{\nu\sigma}+\eta_{\mu\sigma}\eta_{\nu\rho}-\frac14 \eta_{\mu\nu}\eta_{\rho\sigma}\right)\Delta(x), \\
\mbox{R-R field:} && \Delta_{\mu_0\cdots\mu_p;\nu_0\cdots\nu_p}(x) :=\ \sum_{\sigma\in {\cal S}_{p+1}}{\rm sgn}(\sigma)\,\eta_{\mu_0\nu_{\sigma(0)}}\cdots\eta_{\mu_p\nu_{\sigma(p)}}\Delta(x), 
\end{eqnarray}
 where the target space indices run over $0,1\cdots,9$. 


\vspace{5mm}

We then specify how these supergravity fields are coupled to D-branes. 
Suppose that a D$p$-brane is embedded in the target space as 
\begin{equation}
X^\mu=X^\mu(\zeta)
\end{equation}
 where $\zeta^\alpha$ are the worldvolume coordinates with $\alpha=0,1,\cdots,p$. 
The interaction vertices of the D$p$-brane with the supergravity fields
 can be read off from the DBI action with CS term
\begin{equation}
S_{\rm DBI+CS}=T_p\int d^{p+1}\zeta\left[ e^{\frac14(p-3)\Phi} \sqrt{-\hat g}+\hat C_{p+1}\right], 
\end{equation} 
where $T_p$ is the tension of a D$p$-brane and
\begin{equation}
\hat g_{\alpha \beta}=\partial_\alpha X^\mu \partial_\beta X^\nu g_{\mu\nu},
\hspace{1cm}
\hat C^{(p+1)}_{\alpha_1 \cdots \alpha_{p+1}}
=\partial_{\alpha_1} X^{\mu_1}\cdots\partial_{\alpha_{p+1}} X^{\mu_{p+1}} C^{(p+1)}_{\mu_1 \cdots \mu_{p+1}}
\end{equation} 
 are the induced fields on the worldvolume. 
The vertices can be read off from the variations of this action. 
The relevant terms  are
\begin{eqnarray}
\mbox{dilaton:}
&& \frac{p-3}{4}T_p\int d^{p+1}\zeta\,\sqrt{-\det\hat{\eta}_{\alpha\beta}} \, \delta\phi,\\
 [1mm] \mbox{graviton:}
&& -\frac{1}{2}T_p\int d^{p+1}\zeta\,\sqrt{-\det\hat{\eta}_{\alpha\beta}}\,
\hat{\eta}^{\gamma\delta}\partial_{\gamma} X^\mu\partial_{\delta} X^\nu \, \delta g_{\mu\nu}, \\
 [1mm] \mbox{R-R field:}
&& \frac{T_p}{(p+1)!}\int d^{p+1}\zeta\,\epsilon^{\alpha_0\cdots\alpha_p}
\partial_{\alpha_0}X^{\mu_0}\cdots\partial_{\alpha_p}X^{\mu_p} \, \delta C_{\mu_0 \cdots \mu_p}, 
\end{eqnarray} 
 where 
\begin{equation}
\hat{\eta}_{\alpha\beta}:=\partial_\alpha X^\mu \partial_\beta X^\nu \eta_{\mu\nu}, \hspace{1cm} \hat{\eta}^{\alpha\beta}\hat{\eta}_{\beta\gamma}=\delta^\alpha_\gamma. 
\end{equation}
The dilaton vacuum expectation value is absorbed in the string coupling constant. 

\vspace{5mm}

Using the above propagators and interaction vertices,
 the classical potential is given by a sum of contributions
 of the exchanges of the supergravity fields:
\begin{equation}
\tilde{V}_{c}=
-2\kappa_{10}^2\int d^{p+1}\zeta\int d^{p+1}\tilde{\zeta}\,\Delta(X-\tilde{X})
\left( F_\Phi(X,\tilde{X})+F_g(X,\tilde{X})+F_C(X,\tilde{X}) \right), 
\label{V_C general}
\end{equation}
 where 
\begin{eqnarray}
F_\Phi(X,\tilde{X}) 
&=& \left( \frac{p-3}{4} \right)^2T_p^2
 \sqrt{-\det\hat{\eta}_{\alpha\beta}(X)}\sqrt{-\det\hat{\eta}_{\gamma\delta}(\tilde{X})}, \\ [2mm]
F_g(X,\tilde{X}) 
&=& T_p^2\sqrt{-\det\hat{\eta}_{\alpha\beta}(X)}\sqrt{-\det\hat{\eta}_{\gamma\delta}(\tilde{X})}
\nonumber \\
&&\times \left( -\frac{(p+1)^2}{16}
+\frac12\hat{\eta}^{\alpha\beta}(X)(\partial_\beta X\cdot\partial_\delta\tilde{X})
\hat{\eta}^{\delta\gamma}(\tilde{X})(\partial_\gamma\tilde{X}\cdot\partial_\alpha X) \right),
\nonumber \\ \\
F_C(X,\tilde{X}) 
&=& T_p^2\det(\partial_\alpha X\cdot\partial_\beta\tilde{X}). 
\end{eqnarray}
Here, $F_\Phi(X,\tilde{X})$, $F_g(X,\tilde{X})$ and $F_C(X,\tilde{X})$
 are contributions from the dilaton, graviton and RR-fields, respectively. 
Details of the calculations are given in Appendix \ref{appendix:sugra-general}.

\vspace{5mm}
\subsection{Supergravity potential of Revolving D$p$-branes} \label{SUGRA potential}
\vspace{5mm}

We apply the result of eq.(\ref{V_C general}) to the revolving D$p$-branes. 
The embedding functions $X^\mu$ and $\tilde{X}^\mu$ for the revolving D$p$-branes are given by
\begin{equation}
\begin{array}{lll}
X^\alpha=\zeta^\alpha, \hspace{5mm}
 & X^8=r\cos \omega \zeta^0, \hspace{5mm}
 & X^9=r\sin \omega \zeta^0, \\ [2mm]
\tilde{X}^\alpha=\tilde{\zeta}^\alpha, \hspace{5mm}
 & \tilde{X}^8=-r\cos \omega \tilde{\zeta}^0, \hspace{5mm}
 & \tilde{X}^9=-r\sin \omega \tilde{\zeta}^0. 
\end{array}
\end{equation}
Inserting these functions into eq.(\ref{V_C general})
 and performing some of the integrations, we obtain 
\begin{eqnarray}
\tilde{V}_{c} (2r)&=&
-\kappa_{10}^2T_p^2(4\pi)^{-\frac{10-p}{2}}\frac{v^4}{1+v^2}
\int^\infty_{\tilde{\Lambda}^{-2}} ds \ s^{-\frac{10-p}{2}} \nonumber \\
&&\times \int d\zeta  \exp\left[-\frac1{4s} \left(\zeta^2+2r^2(1+\cos \omega \zeta)\right)\right]
(1+\cos \omega \zeta)^2, 
\label{V_C rev}
\end{eqnarray}
 where $v = r \omega$.  
For details of the calculations, see Appendix \ref{appendix:sugra-revolvingDp}.  
Note that  we have performed the Wick rotation of $\zeta^0$ and $\tilde{\zeta}^0$
so that the integral is well-defined. 
$\omega$ is analytically continued as well.

Following the recipe in section \ref{recipe}, 
the cutoff $\tilde{\Lambda}$ is fixed as follows. 
The suppression factor due to the string tension in the above integrand
 is given by ${\rm exp}(-r^2/s)$.
 The cutoff is chosen at $\tilde{s}=1$ when this factor 
 is expressed as ${\rm exp}(-(2r)^2/(2\pi\alpha' \tilde{s}))$. Thus we take
 $s= \pi \alpha' \tilde{s}/2$ and $s_{\rm cutoff}=\tilde{\Lambda}^{-2} =\pi \alpha'/2$.
 Hence $\tilde{\Lambda}$ needs to be fixed at
 $\tilde{\Lambda}=\sqrt{4/(2\pi\alpha')}=2 m_s$.

Several comments are in order. 
First, let us investigate the large $r$ behavior of the potential with $v$ fixed as a small value. 
The integral eq.(\ref{V_C rev}) becomes
\begin{equation}
\tilde{V}_{c} (2r)=
 -(4\pi^2\alpha')^{3-p}(4\pi)^{-\frac{7-p}{2}}\Gamma({\scriptstyle \frac{7-p}{2}})\frac{v^4}{r^{7-p}}
 +{\cal O}(v^6), 
\label{sugra-large-r}
\end{equation}
It reproduces the effective potential
 for two D$p$-branes moving with the relative velocity $2v$ and the impact parameter $2r$,
which can be calculated in string worldsheet theory
 (see eq.(13.5.7) in \cite{Polchinski:1998rr}). 
This provides a consistency check for our result in eq.(\ref{V_C rev}). 

We note that the potential from the supergravity calculation in eq.(\ref{sugra-large-r})
 is proportional to $v^4/r^{7-p}$. 
This behavior in case of $p=0$ 
 is well-known in the calculation of D0-brane scattering in the BFSS matrix theory
 \cite{Banks:1996vh,Seiberg:1997ad}. 
As mentioned at the end of section \ref{VO},
 the same potential can be reproduced from the SYM calculation,
 if we take the UV cutoff to infinity $\Lambda \rightarrow \infty.$ 
In our calculation,
 $\Lambda$ needs to be fixed at $m_s$ in order to avoid the double counting,
and the behavior of the Newton potential 
at large $r$ is generated only by the supergravity calculation. 

There is no chance of a bound state at large distances $r > l_s$.
The potential is proportional to $-\omega^4 r^{p-3}$ and a very weak attractive potential.
Indeed, if angular momentum of the revolving D-brane is conserved, $\omega$ is proportional to $1/r^2$.
Then the potential is proportional to $- r^{p-11}$. Though it is attractive, 
the attractive force is too weak to balance with the repulsive centrifugal potential
which is proportional to $1/r^2$. 

Finally, note that the potential in eq.(\ref{V_C rev}) is 
proportional to $v^4=\omega^4 r^4$ and 
 the $v^2=\omega^2 r^2$ terms are cancelled. It is 
contrary  to a naive expectation that there are large radiative corrections 
to the $\omega^2 r^2$ term in  the effective potential:  
 the supersymmetry breaking scale is given by $\omega$. 
We come back to this property in the next section.

\vspace{5mm}

\section{One-loop effective potential at all ranges of $r$ \label{Epotential} }

\vspace{5mm}

We now investigate the behavior of the one-loop contributions, in the sense of open-strings, to the effective potential at all ranges of $r$
by adding the contributions from SYM and supergravity;
 $\tilde{V}(2r)= \tilde{V}_o(2r)+\tilde{V}_c(2r)$. 
Here we assume that the angular frequency is small compared to the string scale,
 $\omega \ll m_s$ and  the pair of D$p$-branes are revolving slowly. 
We mainly focus on the $p=3$ case. 
D0-branes are also interesting from the BFSS matrix theory point of view,
 since a threshold bound state is expected to arise \cite{Kabat:1996cu,Danielsson:1996uw}. 
 We leave its detailed analysis for future investigations. 

In the following, we recover $\alpha'$ and the ``distance'' $r$ is defined to have mass dimension 1. 
The gauge theory results turn out to be intact by regarding $r$ as a quantity with mass dimension $1$. 
For the supergravity result, 
 we need to replace $r$ with $2\pi\alpha'r$ in order to combine it
 with the gauge theory result for obtaining the effective potential
in the worldvolume effective field theory. 

The contributions from open light modes to the potential
 in Euclidean signature is given by a sum of these two contributions, 
\begin{eqnarray}
\tilde{V}_{o,B}(2r)&= &
 -\int_{\Lambda^{-2}}^\infty \frac{dt}{t}\int\frac{d^{p+1}k}{(2\pi)^{p+1}} e^{-(k^2+4r^2)t}
\nonumber \\
&& \times 
\left[ 6+2e^{-\omega^2t+\frac{8(r\omega)^2}{k^2+4r^2}t}
 \cosh\left( t\sqrt{4\omega^2k_\tau^2+\left( \frac{8(r\omega)^2}{k^2+4r^2} \right)^2} \right) \right],
\nonumber \\ [2mm]
\tilde{V}_{o,F}(2r)&= &
 8\int_{\Lambda^{-2}}^\infty\frac{dt}{t}\int\frac{d^{p+1}k}{(2\pi)^{p+1}}e^{-(k^2+4r^2)t}e^{-\frac14\omega^2t}
 \cosh\left( t\sqrt{\omega^2k_\tau^2+4(r\omega)^2} \right),
\nonumber \\ [2mm]
\label{openpotential-summary}
\end{eqnarray}
 where the UV cutoff is fixed as $\Lambda=\sqrt{1/2\pi\alpha' }= m_s$.
They are complicated integrals and the behaviors at small $r$ and $\omega$ are nontrivial. 
We first look at some general behaviors. 
First, as discussed at the end of section \ref{VO},
 the potential is exponentially damped $\tilde{V}_o \sim e^{-4r^2/\Lambda^2}$
 at large $r > \Lambda=m_s$. 
In the small $r$ region, it behaves nontrivially,
though the potential vanishes at $r=0$.
This can be seen by setting $r=0$ in eq.(\ref{openpotential-summary}).
\begin{eqnarray}
\tilde{V}_{o,B}(2r)&= & -\int_{\Lambda^{-2}}^\infty \frac{dt}{t}\int\frac{d^{p+1}k}{(2\pi)^{p+1}}e^{-k^2t} 
\left[ 6+2e^{-\omega^2t}\cosh\left(2 \omega k_\tau t \right) \right],
\nonumber \\ [2mm]
\tilde{V}_{o,F}(2r)&= & 8\int_{\Lambda^{-2}}^\infty\frac{dt}{t}\int\frac{d^{p+1}k}{(2\pi)^{p+1}}e^{-k^2t}
e^{-\frac14\omega^2t}\cosh\left(  \omega k_\tau t \right). 
\end{eqnarray}
Then, the $\omega$ dependence in each integral is removed by a shift of $k_\tau$ variable:
 $k_\tau \rightarrow k_\tau \pm \omega$ for the bosonic contribution and 
 $k_\tau \rightarrow k_\tau \pm \omega/2$ for the fermionic contribution.
We see that the bosonic and fermionic contributions are cancelled at $r=0$
 and $\tilde{V}_o(0)=0$.
Thus, the supersymmetry makes the potential non-singular at $r=0$. 
Similarly the potential $\tilde{V}_o(2r)$ vanishes at $\omega=0$. 

The contributions from the supergravity $\tilde{V}_c(2r)$ in eq.(\ref{V_C rev})
 gives the Newton potential at large $r$
 and the threshold corrections to $\tilde{V}_o(2r)$ at small $r$.
We discuss more details later, but
here note that
 the potential is proportional to $v^4=\omega^4 r^4$,
 and there are no terms proportional to $v^2$. 
As discussed in the introduction,
 since the supersymmetry breaking scale is given by $\omega$,
 we may naively expect large threshold corrections proportional to $\omega^2 r^2$
 from open string massive modes. In the present calculations, 
however, they are cancelled between infinitely many modes, and
 no terms like $\omega^2 r^2$ are generated for the moduli field $r$
in the worldvolume effective field theory. 
It might be a stringy effect with infinitely many particles, and could not
occur in ordinary quantum field theories. 
It is amusing
 if a similar mechanism would be applied to
  the hierarchy problem of the Higgs potential in the Standard Model.

\vspace{5mm}
\subsection{Shape of the one-loop contributions of the effective potential}
\vspace{5mm}

In this section, in order to get an overview of the one-loop effective potential $\tilde{V}(2r)$,
we expand the formulae in eq.(\ref{openpotential-summary}) 
with respect to $\omega$ and perform the integrations.
First, we look at $\tilde{V}_o(2r)$. 
From the integral representation of eq.(\ref{openpotential-summary}),
 the expansion turns out to be an expansion with respect to $\omega/r$.
Thus the validity of the following expansion is restricted in the region $\omega < r$. 
This region is important for phenomenological applications \cite{Iso:2015mva,Iso:2018sgx}. 
Details of the calculations are given in appendix \ref{potential omega-expansion}. 
After analytic continuation $\omega \rightarrow -i \omega$,
 we obtain the effective potential for $p=3$
  in the Lorentzian signature up to ${\cal O}(\omega^4)$;
\begin{eqnarray}
\tilde{V}_{o,B}(2r)
&=& -\frac{\Lambda^4}{4\pi^2}
\left[
 \left( 1-\frac{4r^2}{\Lambda^2} \right)e^{-4r^2/\Lambda^2}+\frac{16r^4}{\Lambda^4}E_1(4r^2/\Lambda^2)
\right]
\nonumber \\ [2mm] 
& &-\omega^2
\left[
 \frac{r^2}{\pi^2}e^{-4r^2/\Lambda^2}
 -\left( \frac{r^2}{\pi^2}+\frac{4r^4}{\pi^2\Lambda^2} \right)E_1(4r^2/\Lambda^2)
\right] \nonumber \\ [2mm] 
& & -\omega^4
\left[
 \left( \frac{1}{24\pi^2}+\frac{2r^2}{3\pi^2\Lambda^2}
 +\frac{10r^4}{3\pi^2\Lambda^4} \right)e^{-4r^2/\Lambda^2}
 -\left( \frac{6r^4}{\pi^2\Lambda^4}
 +\frac{40r^6}{3\pi^2\Lambda^6} \right)E_1(4r^2/\Lambda^2)
\right] \nonumber \\ [2mm] 
& &+{\cal O}(\omega^6), \\ [3mm]
\tilde{V}_{o,F}(2r)
&=& \frac{\Lambda^4}{4\pi^2}
\left[
 \left( 1-\frac{4r^2}{\Lambda^2} \right)e^{-4r^2/\Lambda^2}+\frac{16r^4}{\Lambda^4}E_1(4r^2/\Lambda^2)
\right] \nonumber \\ [2mm] 
& &-\omega^2
\left[
 \frac{r^2}{\pi^2}E_1(4r^2/\Lambda^2) \right]
 -\omega^4\left[ \left( \frac1{48\pi^2}
 -\frac{r^2}{12\pi^2\Lambda^2} \right)e^{-4r^2/\Lambda^2}
\right]+{\cal O}(\omega^6),
\label{SYMpotential-w}
\end{eqnarray}
where $E_n(x)$ are the exponential integral functions defined in eq.(\ref{expintegral}),
 whose small $x$ behavior for $n=1$ is given by 
\begin{equation}
E_1(x) = -\gamma - \log x + x -\frac{x^2}{4} + {\cal O}(x^3) .
\end{equation} 
Both of the bosonic and fermionic contributions
 have quartic and quadratic divergences but they are completely cancelled as expected. 
The sum gives the SYM contribution to the effective potential;
\begin{eqnarray}
\tilde{V}_{o}(2r) 
&= &-\frac{\omega^2 r^2 }{\pi^2} 
\left[ e^{-4r^2/m_s^2}
-\left(\frac{4r^2 }{m_s^2} \right)E_1(4r^2/m_s^2)  \right]
\nonumber \\ [2mm] 
& & -\omega^4
\left[
 \left( \frac{1}{16\pi^2}+\frac{7r^2}{12\pi^2m_s^2}
 +\frac{10r^4}{3\pi^2m_s^4} \right)e^{-4r^2/m_s^2} \right.
\nonumber \\ [2mm]
& & \hspace*{1cm}\left.
 -\left( \frac{6r^4}{\pi^2m_s^4}
 +\frac{40r^6}{3\pi^2m_s^6} \right)E_1(4r^2/m_s^2)
\right]+{\cal O}(\omega^6). 
\label{open-potential-wexpansion}
\end{eqnarray}
Here we have replaced $\Lambda$ by $m_s$.
This formula is valid as far as the condition $\omega <r $ is  satisfied.  
At large $r$ it is exponentially damped and the potential is negative
 so that the corresponding force is attractive.
From a general discussion, we saw that the potential vanishes $\tilde{V}_o(0)=0$ at the origin.  
At small $r$ (but $r>\omega$),
 the potential in eq.(\ref{open-potential-wexpansion}) behaves like the inverted harmonic potential, $-\omega^2 r^2/\pi^2$,
 and the corresponding force is repulsive for a fixed $\omega$. 
The next order term proportional to $\omega^4$ seems to give a constant value at $r=0$ and
contradict with the general discussion $\tilde{V}_o(0)=0$.
However, it is simply because $r=0$ at fixed $\omega$ 
is out of the validity region of the $\omega$ expansion in eq.(\ref{open-potential-wexpansion}). 

In the region $r<\omega$,
 we can perform a different approximation of the integral for $\tilde{V}_o(2r)$
 to estimate the shape of the potential. 
We set $r=\beta \omega$ and expand $\tilde{V}_{o}(2r)$ in terms of $\beta$. 
As a result, we obtain 
\begin{eqnarray}
\tilde{V}_o 
&=& \frac{\beta^2\omega^4}{\pi^2}\left( -\frac{m_s^2}{\omega^2}\left( 1-E_2(\omega^2/m_s^2) \right)+\int_{\omega^2/m_s^2}^\infty\frac{dt}{t}e^{-t/4}F({\textstyle \frac12,\frac32;\frac t4}) \right)+{\cal O}(\beta^4). \nonumber \\
\end{eqnarray}
Details of the calculations are given in the appendix \ref{r/w expansion}. 
The leading order behavior with respect to $\omega/m_s$ is the same as the above $\omega$-expansion
\begin{equation}
\tilde{V}_o(2r) \sim - \frac{\omega^2 r^2}{\pi^2} .
   \label{small r}
\end{equation}
Thus, as far as the leading behavior is concerned,
 eq.(\ref{open-potential-wexpansion}) seems to give a good approximation at small $r$.
 
\vspace{5mm}

The contributions from the supergravity calculations in eq.(\ref{V_C rev})
 can be also obtained by the $\omega$ expansion.  
In this case, the expansion is with respect to $v=\omega r$,
 and the validity holds as far as $\omega r < 1$
 (here, the mass dimension of $r$ is taken to be $-1$).
Recall that, in this section, $r$ is defined to have mass dimension $1$.
Thus we need to multiply $r$ in eq.(\ref{V_C rev}) by $1/2\pi\alpha'=m_s^2$.
After expanding eq.(\ref{V_C rev}) with respect to $\omega$, the integrals
can be easily performed and we obtain 
\begin{eqnarray}
\tilde{V}_c(2r)
  &=& -\frac{\omega^4}{16\pi^2}
\left[
 1-\left( 1+4 r^2/m_s^2 \right)e^{-4r^2/m_s^2}
\right] 
+{\cal O}(\omega^6)
\label{closed-potential-wexpansion}
\end{eqnarray}
At large $r$, it is approximated by
\begin{equation}
 \tilde{V}_c(2r) \sim \frac{-\omega^4}{16 \pi^2} = -\frac{v^4}{16\pi^2 r^4}
\end{equation}
 with $v=\omega r$, which  reproduces the Newton-like potential for D3-branes in D=10. 
At small $r$, it becomes 
\begin{equation}
 \tilde{V}_c(2r) \sim -\frac{\omega^4 r^4}{2\pi^2m_s^4 } .
\end{equation}
Note that a naively expected term $v^2= \omega^2 r^2$ is absent and the potential
starts from $v^4$. It has been  known in the large $r$ behavior of the D-brane potential,
but it has also an important  implication in the small $r$ behavior of the effective potential
in the field theory of D-branes.


\vspace{5mm}
 
 \begin{figure}[!htb]
\center
\includegraphics [scale=.8] {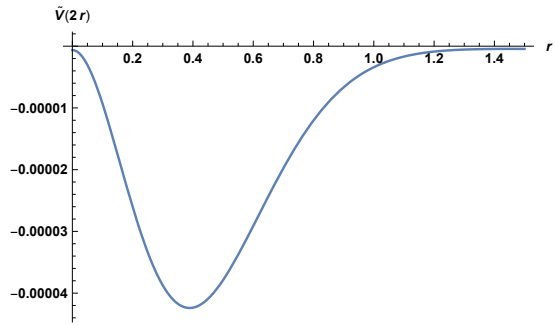}
\caption{\small The shape of the effective potential $\tilde{V}(2r)$
 (the sum of eqs.(\ref{open-potential-wexpansion}) and (\ref{closed-potential-wexpansion}))
 with $\omega=0.1$ and $\Lambda=1$. }
\label{Fig-Vshape}
\end{figure}

Now we sum the contributions from SYM and supergravity. 
The shape of the potential $\tilde{V}(2r)$ is drawn in figure \ref{Fig-Vshape} with $\omega$ fixed at $0.1$. 
At large $r$, Newton potential is reproduced and the corresponding force is attractive.
At small $r$, there is a minimum of the one-loop potential and the corresponding force is repulsive. 
In the intermediate region of $r$, both of the SYM and supergravity contribute to the 
potential. 
In the next section, we briefly argue a possibility of a bound state by combining both of the classical centrifugal potential
and the one-loop effective potential discussed above.

\vspace{5mm}
\subsection{Can the revolving D3-branes form a bound state?}

Finally we briefly argue  whether there exists  a bound state of revolving D3-branes 
with the potential $\tilde{V}(2r)$ studied above. 
Assume that the angular momentum is conserved and there are no quantum radiation. 
We then need to take into account the effect of the centrifugal potential for the D3-branes. 
Also it is necessary to study the behavior of the potential
 with fixing the angular momentum $L$ of the D3-branes per unit volume, instead of the angular frequency $\omega$. 
 
\begin{figure}[!htb]
\center
\includegraphics [scale=.8]
 {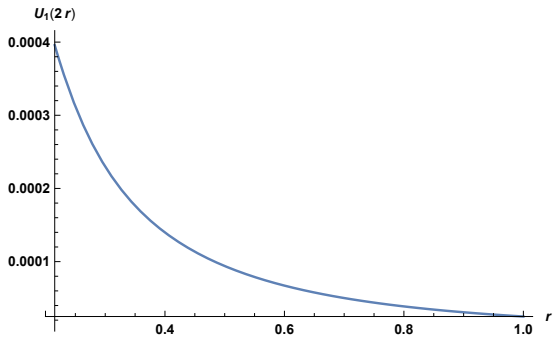}
\caption{\small 
The shape of the potential $U(2r)$ with $T_3=1$, $L=0.01$ and $\Lambda=1$.
}
 \label{Fig-NoBound}
\end{figure}

The potential we need to study is given by 
\begin{equation}
U(2r):=\frac{L^2}{4T_3r^2}+\tilde{V}(2r)
\end{equation}
 with $\omega$ replaced with $L/T_3r^2$. 
The relative distance and reduced mass for a unit volume
 is given by $2r$ and $T_3/2=m_s^4/4\pi g_s$.
Since our calculations are based on the one-loop string calculations,
 the string coupling constant should be smaller than 1. 
In such a situation,
 the potential $U(2r)$ behaves like in Figure \ref{Fig-NoBound}, and there is no minimum,
 because the centrifugal potential is more dominant
 than the induced potential by the one-loop calculations.
It excludes a possibility of forming a bound state for revolving two D3-branes
as long as the string coupling is weak. 

\begin{figure}[!htb]
\center
\includegraphics [scale=.8]
 {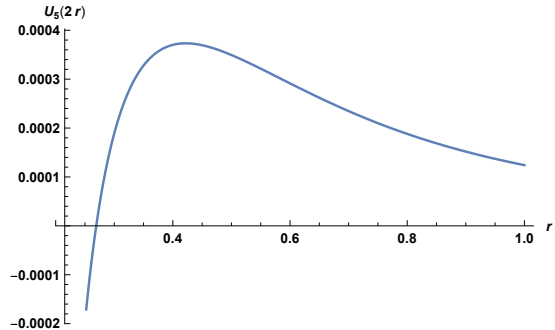}
\caption{\small 
The shape of the potential $U_N(2r)$  with $T_3=1$, $L=0.01$ and $\Lambda=1, N=5$.
}
 \label{Fig-BoundState}
\end{figure}

The situation is changed if we consider a stack of $N$ D3-branes revolving around each other. 
Suppose that each of the revolving D3-branes are replaced with $N$ D3-branes. 
Then, $\tilde{V}(2r)$ is multiplied by $N^2$,
 since there are $N^2$ open strings stretched between the two sets of D3-branes. 
On the other hand, the centrifugal potential is multiplied  by $N$. 
Therefore, the potential $U(2r)$ is modified as 
\begin{equation}
U_N(2r):=\frac{NL^2}{4T_3r^2}+N^2\tilde{V}(2r). 
\end{equation}
For a sufficiently large $N$, 
 the behavior of $U_N(2r)$ changes to the figure drawn in Figure \ref{Fig-BoundState},
 which is qualitatively different from $U(2r)$. 
The potential at small $r$ in eq.(\ref{small r}) shows that 
the potential $U_N(r)$ falls off as $r^{-2}$ for small $r$ after replacing 
$\omega$ by $1/r^2$. 
It is still questionable if a stable bound state exists, but it is
amusing that the potential shows different behavior at small $r$. 

\section{Conclusions and Discussions} \label{discussion}

In this paper,
 we have calculated the one-loop effective potential between  revolving parallel D$p$-branes (especially $p=3$)
 in ten-dimensional space-time. 
Since the end points of open strings attached to the D$p$-branes are accelerating, 
 the boundary conditions  become complicated,
 and we cannot exactly obtain the spectrum of open strings. 
Thus, in the usual method with string worldsheet theory,
 it is difficult to calculate one-loop open string amplitudes
 to obtain the shape of the potential in the whole ranges of $r$. 
 
In this paper, 
 we have introduced a method of {\it partial modular transformation}
 and calculated the effective potential 
 without resorting to the conventional method to obtain open string amplitudes. 
Our method of the partial modular transformation 
 is to perform the modular transformation only in the UV region of the open string modular parameter.
Then, the one-loop open string amplitudes can be approximated
 by a sum of the one-loop amplitudes in the SYM effective worldvolume field theory 
and tree-level amplitudes in supergravity theory.
Corresponding to the partial modular transformation,
 appropriate cutoffs in both theories are introduced, 
 which can remove undesirable double counting of open and closed channels.
 Furthermore, the approximation is good with an accuracy of less than $3 \%$.

We then applied the method
 to a system of  revolving parallel D$p$-branes 
 (in particular, D3-branes) with angular frequency $\omega$ and the relative distance $2r$. 
In the SYM side, 
 we calculated one-loop field theory amplitudes in a background
 corresponding to the revolving motion of D-branes. 
In the supergravity side,
 we calculated potential generated by exchanges of supergravity fields between revolving D-branes. 
Summing these contributions with appropriate cutoffs,
 we have obtained a potential $\tilde{V}(2r)$ in the whole ranges of $r$. 
At large $r$, the supergravity potential is reproduced. 
At small $r$, we found that the potential has a minimum for a fixed $\omega$,
 and the whole shape is drawn in Figure \ref{Fig-Vshape}.

From the field theory point of view,
 $r$ is the moduli field in effective worldvolume SYM field theory.
Due to the supersymmetry, the moduli field is massless at $\omega=0$
and expected to acquire a mass proportional to $\omega$ by supersymmetry breaking. 
It is indeed the case in the SYM calculations;
 the $\omega^2 r^2$ term is radiatively generated in the effective potential. 
On the other hand, 
the supergravity calculation, 
 which corresponds to one-loop amplitudes of open string massive states,
 shows that 
 the leading order term of the potential is given by ${\cal O}(\omega^4)\,$\footnote{
 This fact is known in the context of D0-brane scattering at large $r$, e.g. \cite{Douglas:1996yp}.}, 
 and no terms like $\omega^2 r^2$ arise. 
The cancellation of the threshold corrections to the
 term $\omega^2 r^2$ among infinitely many massive modes 
will be related to the large supersymmetries in the bulk space-time. 
We hope to apply the cancellation mechanism of the stringy threshold corrections
 to the hierarchy problem of the Higgs potential.
 
 Another interesting behavior is the shape of the effective potential and a possibility 
 of a bound state. The potential in Figure \ref{Fig-Vshape} has a minimum
 in small $r$ region, but when we discuss a bound state, we need to take 
 the centrifugal potential into account.  
 With the angular momentum kept fixed, the shape of the potential is changed to
 Figure \ref{Fig-NoBound} and the minimum disappears. 
 The balance between the centrifugal potential and the induced effective potential is, however,  
 subtle and if we consider a stack of N D-branes, the shape of the potential might change to
 Figure \ref{Fig-BoundState}. 
Then the next task is to quantize the collective coordinate, i.e. D-brane relative motion, in 
 the potential of Figure \ref{Fig-BoundState}.  
 In future investigations,
 we want to come back to the issue of bound states 
and to construct phenomenologically viable models.

\vspace{1cm}

{\Large \bf Acknowledgements}

\vspace{3mm}

We would like to thank Takeshi Morita for enlightening discussions.
This work is supported in part by Grants-in-Aid for Scientific Research
 No.~16K05329, No.~18H03708 and No.~19K03851
 from the Japan Society for the Promotion of Science. 

\vspace{1cm}

\appendix

\section{Supergravity potential between D$p$-branes}\label{SUGRA}

\vspace{5mm}

In this appendix
 we calculate the classical potential between revolving D$p$-branes
 induced by exchange of massless supergravity fields. 
First, consider a general configuration of a pair of D$p$-branes. 
Their trajectories $X^\mu(\zeta)$ and $\tilde{X}^\mu(\tilde{\zeta})$ are arbitrary. 
The propagators and the interaction vertices necessary for the following calculations
 are given in subsection \ref{propagator}. 

\vspace{5mm}
\subsection{General formula for supergravity potential} \label{appendix:sugra-general}
\vspace{5mm}

The dilaton exchange gives a contribution to the potential as
\begin{equation}
-\left( \frac{p-3}{4} \right)^2T_p^2\int d^{p+1}\zeta\int d^{p+1}\tilde{\zeta}\,\sqrt{-\det\hat{\eta}_{\alpha\beta}(X)}\sqrt{-\det\hat{\eta}_{\gamma\delta}(\tilde{X})}\Delta(X-\tilde{X}). 
\end{equation}

The graviton exchange gives a contribution to the potential as
\begin{eqnarray}
&& -\frac14T_p^2\int d^{p+1}\zeta\int d^{p+1}\tilde{\zeta}\sqrt{-\det\hat{\eta}_{\alpha\beta}(X)}\sqrt{-\det\hat{\eta}_{\gamma\delta}(\tilde{X})}\,  \nonumber \\ [1mm]
&& \hspace*{1cm}\times \hat{\eta}^{\alpha\beta}(X)\partial_\alpha X^\mu\partial_\beta X^\nu\hat{\eta}^{\gamma\delta}(\tilde{X})\partial_\gamma \tilde{X}^\rho\partial_\delta X^\sigma\Delta_{\mu\nu,\rho\sigma}(X-\tilde{X}). 
\end{eqnarray}
The integrand can be simplified as follows: 
\begin{eqnarray}
& & \hat{\eta}^{\alpha\beta}(X)\partial_\alpha X^\mu\partial_\beta X^\nu\hat{\eta}^{\gamma\delta}(\tilde{X})\partial_\gamma \tilde{X}^\rho\partial_\delta \tilde{X}^\sigma\Delta_{\mu\nu,\rho\sigma}(X-\tilde{X}) \nonumber \\
&=& \hat{\eta}^{\alpha\beta}(X)\hat{\eta}^{\gamma\delta}(\tilde{X})\partial_\alpha X^\mu\partial_\beta X^\nu\partial_\gamma \tilde{X}^\rho\partial_\delta \tilde{X}^\sigma\left( \eta_{\mu\rho}\eta_{\nu\sigma}+\eta_{\mu\sigma}\eta_{\nu\rho}-\frac14\eta_{\mu\nu}\eta_{\rho\sigma} \right)\Delta(X-\tilde{X}) \nonumber \\
&=& \hat{\eta}^{\alpha\beta}(X)\hat{\eta}^{\gamma\delta}(\tilde{X})\left( 2(\partial_\alpha X\cdot\partial_\gamma \tilde{X})(\partial_\beta X\cdot\partial_\delta \tilde{X})-\frac14(\partial_\alpha X\cdot\partial_\beta X)(\partial_\gamma\tilde{X}\cdot\partial_\delta\tilde{X}) \right)\Delta(X-\tilde{X}) \nonumber \\
&=& \left( 2\hat{\eta}^{\alpha\beta}(X)(\partial_\beta X\cdot\partial_\delta\tilde{X})\hat{\eta}^{\delta\gamma}(\tilde{X})(\partial_\gamma\tilde{X}\cdot\partial_\alpha X)-\frac{(p+1)^2}{4} \right)\Delta(X-\tilde{X}). 
\end{eqnarray}
Therefore, we obtain 
\begin{eqnarray}
& & \frac{(p+1)^2}{16}T_p^2\int d^{p+1}\zeta\int d^{p+1}\tilde{\zeta}\,\sqrt{-\det\hat{\eta}_{\alpha\beta}(X)}\sqrt{-\det\hat{\eta}_{\gamma\delta}(\tilde{X})}\,\Delta(X-\tilde{X}) \nonumber \\
& & -\frac12T_p^2\int d^{p+1}\zeta\int d^{p+1}\tilde{\zeta}\,\sqrt{-\det\hat{\eta}_{\alpha\beta}(X)}\sqrt{-\det\hat{\eta}_{\gamma\delta}(\tilde{X})}\,  \nonumber \\ [1mm]
&& \hspace*{1cm}\times \hat{\eta}^{\alpha\beta}(X)(\partial_\beta X\cdot\partial_\delta\tilde{X})\hat{\eta}^{\delta\gamma}(\tilde{X})(\partial_\gamma\tilde{X}\cdot\partial_\alpha X)\Delta(X-\tilde{X}). 
\end{eqnarray}


The R-R field exchange gives a contribution to the potential as
\begin{eqnarray}
& & -\frac{T_p^2}{((p+1)!)^2}\int d^{p+1}\zeta\int d^{p+1}\tilde{\zeta}\,\epsilon^{\alpha_0\cdots\alpha_p}\partial_{\alpha_0}X^{\mu_0}\cdots\partial_{\alpha_p}X^{\mu_p}
\epsilon^{\beta_0\cdots\beta_p}\partial_{\beta_0}\tilde{X}^{\nu_0}\cdots\partial_{\beta_p}\tilde{X}^{\nu_p} \nonumber \\ [1mm]
& & \hspace*{1cm}\times\eta_{\mu_0\cdots\mu_p;\nu_0\cdots\nu_p}\Delta(X-\tilde{X}), 
\end{eqnarray}
 where 
\begin{equation}
\eta_{\mu_0\cdots\mu_p;\nu_0\cdots\nu_p}\ :=\ \sum_{\sigma\in {\cal S}_{p+1}}{\rm sgn}(\sigma)\,\eta_{\mu_0\nu_{\sigma(0)}}\cdots\eta_{\mu_p\nu_{\sigma(p)}}. 
\end{equation}
The integrand can be simplified as follows: 
\begin{eqnarray}
& & \epsilon^{\alpha_0\cdots\alpha_p}\partial_{\alpha_0}X^{\mu_0}\cdots\partial_{\alpha_p}X^{\mu_p}
\epsilon^{\beta_0\cdots\beta_p}\partial_{\beta_0}\tilde{X}^{\nu_0}\cdots\partial_{\beta_p}\tilde{X}^{\nu_p}
\eta_{\mu_0\cdots\mu_p,\nu_0\cdots\nu_p} \nonumber \\
&=& \sum_{\sigma\in{\cal S}_{p+1}}{\rm sgn}\,(\sigma)\epsilon^{\alpha_0\cdots\alpha_p}\partial_{\alpha_0}X^{\mu_0}\cdots\partial_{\alpha_p}X^{\mu_p}
\epsilon^{\beta_0\cdots\beta_p}\partial_{\beta_0}\tilde{X}^{\nu_0}\cdots\partial_{\beta_p}\tilde{X}^{\nu_p}
\eta_{\mu_0\nu_{\sigma(0)}}\cdots\eta_{\mu_p\nu_{\sigma(p)}} \nonumber \\
&=& \sum_{\sigma\in{\cal S}_{p+1}}{\rm sgn}\,(\sigma)\epsilon^{\alpha_0\cdots\alpha_p}\partial_{\alpha_0}X^{\mu_0}\cdots\partial_{\alpha_p}X^{\mu_p}
\nonumber \\
& & \hspace*{1cm}\times{\rm sgn}(\sigma)\,\epsilon^{\beta_{\sigma(0)}\cdots\beta_{\sigma(p)}}\partial_{\beta_{\sigma(0)}}\tilde{X}^{\nu_{\sigma(0)}}\cdots
\partial_{\beta_{\sigma(p)}}\tilde{X}^{\nu_{\sigma(p)}}\eta_{\mu_0\nu_{\sigma(0)}}\cdots
\eta_{\mu_p\nu_{\sigma(p)}} \nonumber \\
&=& \sum_{\sigma\in{\cal S}_{p+1}}\epsilon^{\alpha_0\cdots\alpha_p}\epsilon^{\beta_0\cdots\beta_p}\partial_{\alpha_0}X^{\mu_0}
\cdots\partial_{\alpha_p}X^{\mu_p}\partial_{\beta_0}\tilde{X}^{\nu_0}\cdots\partial_{\beta_p}\tilde{X}^{\nu_p}
\eta_{\mu_0\nu_0}\cdots\eta_{\mu_p\nu_p} \nonumber \\
&=& (p+1)!\,\epsilon^{\alpha_0\cdots\alpha_p}\epsilon^{\beta_0\cdots\beta_p}(\partial_{\alpha_0}X\cdot\partial_{\beta_0}\tilde{X})\cdots(\partial_{\alpha_p}X\cdot\partial_{\beta_p}\tilde{X}) \nonumber \\
&=& ((p+1)!)^2\det(\partial_\alpha X\cdot\partial_\beta\tilde{X}). 
\end{eqnarray}
Therefore, we obtain 
\begin{equation}
-T_p^2\int d^{p+1}\zeta\int d^{p+1}\tilde{\zeta}\,\det(\partial_\alpha X\cdot\partial_\beta\tilde{X})\,\Delta(X-\tilde{X}). 
\end{equation}


In summary, the supergravity potential is totally given by
\begin{equation}
-2\kappa_{10}^2\int d^{p+1}\zeta\int d^{p+1}\tilde{\zeta}\,\Delta(X-\tilde{X})\left( F_\Phi(X,\tilde{X})+F_g(X,\tilde{X})+F_C(X,\tilde{X}) \right), 
\end{equation}
 where 
\begin{eqnarray}
F_\Phi(\zeta,\tilde{\zeta}) 
&=& \left( \frac{p-3}{4} \right)^2T_p^2\sqrt{-\det\hat{\eta}_{\alpha\beta}(X)}\sqrt{-\det\hat{\eta}_{\gamma\delta}(\tilde{X})}, \\
F_g(\zeta,\tilde{\zeta}) 
&=& T_p^2\sqrt{-\det\hat{\eta}_{\alpha\beta}(X)}\sqrt{-\det\hat{\eta}_{\gamma\delta}(\tilde{X})}\left( -\frac{(p+1)^2}{16} \right. \nonumber \\
& & \hspace*{1cm}\left. +\frac12\hat{\eta}^{\alpha\beta}(X)(\partial_\beta X\cdot\partial_\delta\tilde{X})\hat{\eta}^{\delta\gamma}(\tilde{X})(\partial_\gamma\tilde{X}\cdot\partial_\alpha X) \right), \\
F_C(\zeta,\tilde{\zeta}) 
&=& T_p^2\det(\partial_\alpha X\cdot\partial_\beta\tilde{X}). 
\end{eqnarray}

\vspace{5mm}
\subsection{Supergravity potential in D1-branes at angle}
\vspace{5mm}

As a simple check of our formula in appendix \ref{SUGRA}, 
let us consider a simple example of D1-branes at angle. 
Their trajectories are given by
\begin{eqnarray}
&& X^\alpha\ =\ \zeta^\alpha, \hspace{5mm} (\alpha=0,1) \nonumber \\
&& \tilde{X}^0\ =\ \tilde{\zeta}^0, \hspace{5mm} \tilde{X}^1\ =\ \cos\phi\,\tilde{\zeta}^1, \hspace{5mm} \tilde{X}^2\ =\ \sin\phi\,\tilde{\zeta}^1, \hspace{5mm} \tilde{X}^9\ =\ r, 
\end{eqnarray}
 and zero otherwise. 
It is easy to find that 
\begin{equation}
\hat{\eta}_{\alpha\beta}(X)\ =\ \hat{\eta}_{\alpha\beta}(\tilde{X})\ =\ \eta_{\alpha\beta}, \hspace{1cm} \partial_\alpha X\cdot\partial_\beta\tilde{X}\ =\ \left[ 
\begin{array}{cc}
-1 & 0 \\
0 & \cos\phi
\end{array}
\right]. 
\end{equation}
Then, we obtain 
\begin{eqnarray}
& & \hat{\eta}^{\alpha\beta}(X)(\partial_\beta X\cdot\partial_\delta\tilde{X})\hat{\eta}^{\delta\gamma}(\tilde{X})(\partial_\gamma\tilde{X}\cdot\partial_\alpha X) \nonumber \\
&=& {\rm Tr}\left[
\begin{array}{cc}
-1 & 0 \\
0 & 1
\end{array}
\right]\left[
\begin{array}{cc}
-1 & 0 \\
0 & \cos\phi
\end{array}
\right]\left[
\begin{array}{cc}
-1 & 0 \\
0 & 1
\end{array}
\right]\left[
\begin{array}{cc}
-1 & 0 \\
0 & \cos\phi
\end{array}
\right] \nonumber \\
&=& 1+\cos^2\phi, 
\end{eqnarray}
 and 
\begin{equation}
\det(\partial_\alpha X\cdot\partial_\beta\tilde{X})\ =\ -\cos\phi. 
\end{equation}
Therefore, 
\begin{eqnarray}
& & F_\Phi(X,\tilde{X})+F_g(X,\tilde{X})+F_C(X,\tilde{X}) \nonumber \\
&=& -\left( \frac{1-3}{4} \right)^2T_p^2-\left( -\frac{(1+1)^2}{16}+\frac{1+\cos^2\phi}{2} \right)T_p^2+\cos\phi\,\rho_p^2 \nonumber \\
&=& -2T_1^2\sin^2\frac\phi2. 
\end{eqnarray}
This reproduces the known result \cite{Polchinski:1998rr}. 

\vspace{5mm}

\subsection{Supergravity potential between revolving branes} \label{appendix:sugra-revolvingDp}

\vspace{5mm}

The trajectories are given as 
\begin{equation}
\begin{array}{lll}
X^\alpha=\zeta^\alpha, \hspace{5mm} & X^8=r\cos \omega \zeta^0, \hspace{5mm} & X^9=r\sin \omega \zeta^0, \\ [2mm]
\tilde{X}^\alpha=\tilde{\zeta}^\alpha, \hspace{5mm} & \tilde{X}^8=-r\cos \omega \tilde{\zeta}^0, \hspace{5mm} & \tilde{X}^9=-r\sin \omega \tilde{\zeta}^0, 
\end{array}
\end{equation}
where $\alpha=0,1,\cdots,p$, and $X^\mu,\tilde{X}^\mu=0$ otherwise.
We obtain 
\begin{eqnarray}
\hat{\eta}_{\alpha\beta}(X)\ =\ \left[
\begin{array}{cc}
-1+v^2 & 0 \\
0 & {\bf 1}_p
\end{array}
\right]\ =\ \hat{\eta}_{\alpha\beta}(\tilde{X}). 
\end{eqnarray}
Therefore, 
\begin{eqnarray}
\hat{\eta}^{\alpha\beta}(X)\ =\ \left[
\begin{array}{cc}
(-1+v^2)^{-1} & 0 \\
0 & {\bf 1}_p
\end{array}
\right]\ =\ \hat{\eta}^{\alpha\beta}(\tilde{X}), 
\end{eqnarray}
 and 
\begin{equation}
\sqrt{-\det\hat{\eta}_{\alpha\beta}(X)}\ =\ \sqrt{1-v^2}\ =\ \sqrt{-\det\hat{\eta}_{\alpha\beta}(\tilde{X})}. 
\end{equation}
We also obtain 
\begin{equation}
\partial_\alpha X\cdot\partial_\beta\tilde{X}\ =\ \left[
\begin{array}{cc}
-1-v^2\cos\omega(\zeta^0-\tilde{\zeta}^0) & 0 \\
0 & {\bf 1}_p
\end{array}
\right]. 
\end{equation}
Then 
\begin{eqnarray}
\hat{\eta}^{\alpha\beta}(X)(\partial_\beta X\cdot\partial_\delta\tilde{X})\hat{\eta}^{\delta\gamma}(\tilde{X})(\partial_\gamma\tilde{X}\cdot\partial_\alpha X) 
&=& \frac{(1+v^2\cos\omega(\zeta^0-\tilde{\zeta}^0))^2}{(1-v^2)^2}+p, \nonumber \\ \\
\det(\partial_\alpha X\cdot\partial_\beta\tilde{X}) 
&=& -(1+v^2\cos\omega(\zeta^0-\tilde{\zeta}^0)). 
\end{eqnarray}

Now, we find 
\begin{eqnarray}
F_\Phi(X,\tilde{X}) 
&=& T_p^2\left( \frac{p-3}{4} \right)^2(1-v^2), \\
F_g(X,\tilde{X}) 
&=& T_p^2\left\{ -\frac{(p+1)^2}{16}(1-v^2)+\frac12\left[ \frac{(1+v^2\cos\omega(\zeta^2-\tilde{\zeta}^0))^2}{1-v^2}+p(1-v^2) \right] \right\} \nonumber \\
\\
F_C(X,\tilde{X}) 
&=& -\rho_p^2(1+v^2\cos\omega(\zeta^2-\tilde{\zeta}^0)). 
\end{eqnarray}
They give 
\begin{equation}
F_\Phi(X,\tilde{X})+F_g(X,\tilde{X})+F_C(X,\tilde{X})\ =\ T_p^2\frac{v^4}{2(1-v^2)}\left( 1+\cos\omega(\zeta^0-\tilde{\zeta}^0) \right)^2. 
\end{equation}
Then, the effective potential becomes
\begin{eqnarray}
& & -2\kappa^2\int d^{p+1}\zeta\int d^{p+1}\tilde{\zeta}\,\Delta(X-\tilde{X})\left( F_\Phi(X,\tilde{X})+F_g(X,\tilde{X})+F_C(X,\tilde{X}) \right) \nonumber \\
&=& -\kappa^2T_p^2\frac{v^4}{1-v^2}\int d^{p+1}\zeta\int d^{p+1}\tilde{\zeta}\,\Delta(X-\tilde{X})\left( 1+\cos\omega(\zeta^0-\tilde{\zeta}^0) \right)^2. 
\end{eqnarray}
The integral can be rewritten as follows. 
\begin{eqnarray}
& & \int d^{p+1}\zeta\int d^{p+1}\tilde{\zeta}\,\Delta(X-\tilde{X})\left( 1+\cos\omega(\zeta^0-\tilde{\zeta}^0) \right)^2 \nonumber \\
&=& V_p\int d\zeta^0\int d\tilde{\zeta}^0\int\frac{d^{10-p}k}{(2\pi)^{10-p}}\frac1{k^2}\left( 1+\cos\omega(\zeta^0-\tilde{\zeta}^0) \right)^2 \nonumber \\
& & \hspace*{1cm}\times \exp\left( {ik_\tau(\zeta^0-\tilde{\zeta}^0)+ik_9r(\cos\omega\zeta^0+\cos\omega\tilde{\zeta}^0)
+ik_9r(\sin\omega\zeta^0+\sin\omega\tilde{\zeta}^0)} \right) \nonumber \\
&=& V_p\int d\zeta^0\int d\tilde{\zeta}^0(4\pi)^{-\frac{10-p}2}\int_0^\infty ds\,s^{-\frac{10-p}2}\left( 1+\cos\omega(\zeta^0-\tilde{\zeta}^0) \right)^2 \nonumber \\
& & \hspace*{1cm}\times \exp\left( {-\frac1{4s}\left[ -(\zeta^0-\tilde{\zeta}^0)^2+r^2(2+2\cos\omega(\zeta^0-\tilde{\zeta}^0) \right]} \right) \nonumber \\
&=& V_{p+1}(4\pi)^{-\frac{10-p}2}\int d\zeta\int_0^\infty ds\,s^{-\frac{10-p}2}e^{-\frac1{4s}\left[ -\zeta^2+2r^2(1+\cos\omega\zeta) \right]}\left( 1+\cos\omega\zeta \right)^2, 
\end{eqnarray}
where $\zeta:=\zeta^0-\tilde{\zeta}^0$. 
To make this integral well-defined,
 we perform the Wick rotation $\zeta\to-i\zeta$ and the analytic continuation $\omega\to i\omega$. 
The result is given in (\ref{V_C rev}) in section \ref{SUGRA potential}.

\vspace{1cm}

\section{$\omega$ expansion of SYM potential $\tilde{V}_o(r)$}
\label{potential omega-expansion}

\vspace{5mm}

In this appendix, we evaluate the SYM potential by expanding it with respect to $\omega/r$.
Thus its validity is restricted to $\omega<r$. 
The contributions to the effective potential from bosons and the ghost are 
\begin{eqnarray}
\tilde{V}_{o,B}&= & -\int_{\Lambda^{-2}}^\infty \frac{dt}{t}\int\frac{d^{p+1}k}{(2\pi)^{p+1}}e^{-t(k^2+4r^2)} \nonumber \\ [1mm]
& &\hspace*{1cm}\times \left[ 6+2e^{-t\omega^2+t\frac{8(r\omega)^2}{k^2+4r^2}}\cosh\left( t\sqrt{4\omega^2k_\tau^2+\left( \frac{8(r\omega)^2}{k^2+4r^2} \right)^2} \right) \right] \nonumber \\ [2mm]
&=& \int_{\Lambda^{-2}}^\infty\frac{dt}{t}\int\frac{d^{p+1}k}{(2\pi)^{p+1}}e^{-t(k^2+4r^2)}\left[ -8+\omega^2\left( 2t-4k_\tau^2t^2-\frac{16r^2t}{k^2+4r^2} \right) \right. \nonumber \\ [2mm]
& & \left. +\omega^4\left( -t^2+4k_\tau^2t^3-\frac43k_\tau^4t^4-\frac{32k_\tau^2r^2t^3-16r^2t^2}{k^2+4r^2}-\frac{128r^4t^2}{(k^2+4r^2)^2} \right) \right] \nonumber \\ [2mm]
& & +{\cal O}(\omega^6)
\end{eqnarray}
Those from fermions are 
\begin{eqnarray}
\tilde{V}_{o,F}&= & 4\int_{\Lambda^{-2}}^\infty\frac{dt}{t}\int\frac{d^{p+1}k}{(2\pi)^{p+1}}e^{-t(k^2+4r^2)}e^{-t\cdot\frac{\omega^2}{4}}\cdot 2\cosh\left( t\sqrt{\omega^2k_\tau^2+4(r\omega)^2} \right) \nonumber \\ [2mm]
&=& \int_{\Lambda^{-2}}^\infty\frac{dt}{t}\int\frac{d^{p+1}k}{(2\pi)^{p+1}}e^{-t(k^2+4r^2)}\left[ 8+\omega^2\left( -2t+4k_\tau^2t^2+16r^2t^2 \right) \frac{}{} \right. \nonumber \\ [2mm]
& & \left. +\omega^4\left( \frac14t^2-k_\tau^2t^3-4r^2t^3+\frac13k_\tau^4t^4+\frac83k_\tau^2r^2t^4+\frac{16}{3}r^4t^4 \right) \right] \nonumber \\ [2mm]
& & +{\cal O}(\omega^6).  
\end{eqnarray}
In the following, we drop the ${\cal O}(\omega^0)$ terms since they trivially cancel between bosons and fermions. 
For the other terms, the $t$-integration can be done easily. 

The bosonic contribution becomes 
\begin{eqnarray}
& & 
\int\frac{d^{p+1}k}{(2\pi)^{p+1}}e^{-(k^2+4r^2)/\Lambda^2}\left[ \omega^2\left( \frac{2\Lambda^2-4k_\tau^2}{\Lambda^2(k^2+4r^2)}-\frac{16r^2+4k_\tau^2}{(k^2+4r^2)} \right) \right. \nonumber \\ [2mm] 
& & +\omega^4\left( -\frac{3\Lambda^4-12\Lambda^2k_\tau^2+4k_\tau^4}{3\Lambda^6(k^2+4r^2)} +\frac{16\Lambda^2r^2-32k_\tau^2r^2-\Lambda^4+8k_\tau^2\Lambda^2-4k_\tau^4}{\Lambda^4(k^2+4r^2)^2} \right. \nonumber \\ [2mm] 
& & \left. \left. -\frac{128r^4-16\Lambda^2r^2+64k_\tau^2r^2-8k_\tau^2\Lambda^2+8k_\tau^4}{\Lambda^2(k^2+4r^2)^3}-\frac{128r^4+64k_\tau^2r^2+8k_\tau^4}{(k^2+4r^2)^4} \right) \right] \nonumber \\
& & +{\cal O}(\omega^6). 
\end{eqnarray}
The fermionic contribution becomes 
\begin{eqnarray}
& & \int\frac{d^{p+1}k}{(2\pi)^{p+1}}e^{-(k^2+4r^2)/\Lambda^2}\left[ \omega^2\left( \frac{16r^2-2\Lambda^2+4k_\tau^2}{\Lambda^2(k^2+4r^2)}+\frac{16r^2+4k_\tau^2}{(k^2+4r^2)^2} \right) \right. \nonumber \\ [2mm] 
& & \left. +\omega^4\left( \frac{64r^4-48\Lambda^2r^2+32k_\tau^2r^2+3\Lambda^4-12k_\tau^2\Lambda^2+4k_\tau^4}{12\Lambda^6(k^2+4r^2)} \right. \right. \nonumber \\ [2mm] 
& & +\frac{64r^4-32\Lambda^2r^2+32k_\tau^2r^2+\Lambda^4-8k_\tau^2\Lambda^2+4k_\tau^4}{4\Lambda^4(k^2+4r^2)^2} \nonumber \\ [2mm] 
& & \left. \left. +\frac{32r^4-8\Lambda^2r^2+16k_\tau^2r^2-2k_\tau^2\Lambda^2+2k_\tau^4}{\Lambda^2(k^2+4r^2)^3}+\frac{32r^4+16k_\tau^2r^2+2k_\tau^4}{(k^2+4r^2)^4} \right) \right] \nonumber \\
& &+{\cal O}(\omega^6). 
\end{eqnarray}

In the following, we focus on $p=3$. 
By the rotational symmetry, $k_\tau^2$ in the integrand can be replaced with $\frac14k^2$. 
To deal with $k_\tau^4$, we employ the polar coordinates for the momentum. 
Then 
\begin{eqnarray}
\int\frac{d^4k}{(2\pi)^4}f(k^2)k_\tau^4 
&=& \frac1{(2\pi)^4}\int_0^\infty d\kappa \,\kappa^3 f(\kappa^2)\kappa^4\cdot 4\pi\int_0^\pi d\theta\sin^2\theta\cos^4\theta \nonumber \\
&=& \frac1{(2\pi)^4}\int_0^\infty d\kappa \,\kappa^3 f(\kappa^2)\kappa^4\cdot\frac{\pi^2}{4} \nonumber \\
&=& \int\frac{d^4k}{(2\pi)^4}f(k^2)\cdot\frac18k^4. 
\end{eqnarray}
Using this rewriting, the bosonic contribution becomes 
\begin{eqnarray}
& & \int\frac{d^4k}{(2\pi)^4}e^{-(k^2+4r^2)/\Lambda^2}\left[ \omega^2\left( -\frac1{\Lambda^2}+\frac{4r^2+\Lambda^2}{\Lambda^2(k^2+4r^2)}-\frac{12r^2}{(k^2+4r^2)^2} \right) \right. \nonumber \\ [2mm] 
& & +\omega^4\left( -\frac{k^2+4r^2}{6\Lambda^6}+\frac{8r^2+3\Lambda^2}{6\Lambda^6}-\frac{8r^4+24\Lambda^2r^2}{3\Lambda^6(k^2+4r^2)} \right. \nonumber \\ [2mm]
& & \hspace*{1cm}\left. \left. +\frac{24r^2}{\Lambda^4(k^2+4r^2)^2}-\frac{80r^4}{\Lambda^2(k^2+4r^2)^3}-\frac{80r^4}{(k^2+4r^2)^4} \right) \right]+{\cal O}(\omega^6). 
\end{eqnarray}
The fermionic contribution becomes 
\begin{eqnarray}
& & \int\frac{d^4k}{(2\pi)^4}e^{-(k^2+4r^2)/\Lambda^2}\left[ \omega^2\left( \frac1{\Lambda^2}+\frac{12r^2-\Lambda^2}{\Lambda^2(k^2+4r^2)}+\frac{12r^2}{(k^2+4r^2)^2} \right) \right. \nonumber \\ [2mm] 
& & +\omega^4\left( \frac{k^2+4r^2}{24\Lambda^6}+\frac{8r^2-3\Lambda^2}{24\Lambda^6}+\frac{10r^4-6\Lambda^2r^2}{3\Lambda^6(k^2+4r^2)} \right. \nonumber \\ [2mm]
& & \hspace*{1cm} \left. \left. +\frac{10r^4-4\Lambda^2r^2}{\Lambda^4(k^2+4r^2)^2}+\frac{20r^4-4\Lambda^2r^2}{\Lambda^2(k^2+4r^2)^3}+\frac{20r^4}{(k^2+4r^2)^4} \right) \right]+{\cal O}(\omega^6). 
\end{eqnarray}

The $k$-integration can be done as follows. 
\begin{eqnarray}
& & \int\frac{d^4k}{(2\pi)^4}e^{-(k^2+4r^2)/\Lambda^2}\frac1{(k^2+4r^2)^n} \nonumber \\
&=& \frac{2\pi^2}{(2\pi)^4}\int_0^\infty d\kappa\,\kappa^3e^{-(\kappa^2+4r^2)/\Lambda^2}\frac1{(\kappa^2+4r^2)^n} \nonumber \\
&=& \frac1{16\pi^2}\int_0^\infty du\,e^{-(u+4r^2)/\Lambda^2}\frac{u}{(u+4r^2)^n} \nonumber \\
&=& \frac1{16\pi^2}\int_{4r^2}^\infty du\,e^{-u/\Lambda^2}\frac{u-4r^2}{u^n} \nonumber \\
&=& \frac{1}{16\pi^2}(4r^2)^{2-n}\int_1^\infty du\,e^{-4r^2u/\Lambda^2}\frac{u-1}{u^n} \nonumber \\
&=& \frac1{16\pi^2}(4r^2)^{2-n}\left( E_{n-1}(4r^2/\Lambda^2)-E_n(4r^2/\Lambda^2) \right), 
\end{eqnarray}
where $E_n(x)$ are defined as
\begin{equation}
E_n(x)\ :=\ \int_1^\infty du\,\frac{e^{-xu}}{u^n}. 
\label{expintegral}
\end{equation}
For $n\le0$, they are elementary functions:
\begin{equation}
E_0(x)\ =\ \frac1xe^{-x}, \hspace{1cm} E_{-1}(x)\ =\ \frac{x+1}{x^2}e^{-x}, \hspace{1cm} E_{-2}(x)\ =\ \frac{x^2+2x+2}{x^3}e^{-x}, 
\end{equation}
etc. 
Note that $E_n(x)$ with $n>1$ satisfy 
\begin{eqnarray}
E_n(x) 
&=& -\frac{e^{-xu}}{(n-1)u^{n-1}}\Big|_1^\infty-\frac{x}{n-1}\int_1^\infty du\,\frac{e^{-xu}}{u^{n-1}} \nonumber \\
&=& \frac{e^{-x}}{n-1}-\frac{x}{n-1}E_{n-1}(x). 
\end{eqnarray}
Using these recursion relations,
 the effective potential can be written in terms of $E_1(x)$ and elementary functions.
The bosonic contribution becomes 
\begin{eqnarray}
& & \omega^2\left[ \frac{r^2}{\pi^2}e^{-4r^2/\Lambda^2}-\left( \frac{r^2}{\pi^2}+\frac{4r^4}{\pi^2\Lambda^2} \right)E_1(4r^2/\Lambda^2) \right] \nonumber \\ [2mm] 
& & +\omega^4\left[ \left( -\frac{1}{24\pi^2}-\frac{2r^2}{3\pi^2\Lambda^2}-\frac{10r^4}{3\pi^2\Lambda^4} \right)e^{-4r^2/\Lambda^2}+\left( \frac{6r^4}{\pi^2\Lambda^4}+\frac{40r^6}{3\pi^2\Lambda^6} \right)E_1(4r^2/\Lambda^2) \right] \nonumber \\ [2mm] 
& & +{\cal O}(\omega^6). 
\end{eqnarray}
The fermionic contribution becomes 
\begin{eqnarray}
& & \omega^2\left[ \frac{r^2}{\pi^2}E_1(4r^2/\Lambda^2) \right]+\omega^4\left[ \left( -\frac1{48\pi^2}+\frac{r^2}{12\pi^2\Lambda^2} \right)e^{-4r^2/\Lambda^2} \right]+{\cal O}(\omega^6). 
\end{eqnarray}
The sum of these two contributions is 
\begin{eqnarray}
& & \omega^2\left[ \frac{r^2}{\pi^2}e^{-4r^2/\Lambda^2}-\frac{4r^2}{\pi^2\Lambda^2}E_1(4r^2/\Lambda^2) \right] \nonumber \\ [2mm] 
& & +\omega^4\left[ \left( -\frac{1}{16\pi^2}-\frac{7r^2}{12\pi^2\Lambda^2}-\frac{10r^4}{3\pi^2\Lambda^4} \right)e^{-4r^2/\Lambda^2}+\left( \frac{6r^4}{\pi^2\Lambda^4}+\frac{40r^6}{3\pi^2\Lambda^6} \right)E_1(4r^2/\Lambda^2) \right] \nonumber \\ [2mm] 
& & +{\cal O}(\omega^6). 
\end{eqnarray}
Performing the analytic continuation of $\omega$, this becomes 
\begin{eqnarray}
& & -\omega^2\left[ \frac{r^2}{\pi^2}e^{-4r^2/\Lambda^2}-\frac{4r^2}{\pi^2\Lambda^2}E_1(4r^2/\Lambda^2) \right] \nonumber \\ [2mm] 
& & -\omega^4\left[ \left( \frac{1}{16\pi^2}+\frac{7r^2}{12\pi^2\Lambda^2}+\frac{10r^4}{3\pi^2\Lambda^4} \right)e^{-4r^2/\Lambda^2}-\left( \frac{6r^4}{\pi^2\Lambda^4}+\frac{40r^6}{3\pi^2\Lambda^6} \right)E_1(4r^2/\Lambda^2) \right] \nonumber \\ [2mm] 
& & +{\cal O}(\omega^6). 
\end{eqnarray}

The $\omega$-independent terms which we have dropped at the beginning are, as noted,
trivially cancelled between bosons and fermions, 
\begin{equation}
\frac{\Lambda^4}{16\pi^2}E_3(4r^2/\Lambda^2)\cdot(-8+8)=0. 
\end{equation}

\section{$r$ expansion of SYM potential $\tilde{V}_o(r)$}
\label{r/w expansion}

In the region $r<\omega$,
 the expansion of the effective potential in Appendix \ref{potential omega-expansion} is no longer valid
 and we need another method to approximate it. 
In this appendix,
 we set $r=\beta \omega$ and approximate the effective potential in terms of $\beta$-expansion. 
 Thus this evaluation of the effective action is valid in the region of $r<\omega$.
The bosonic and fermionic contributions to the effective action for $p=3$ are rewritten as
\begin{eqnarray}
\tilde{V}_{o,B}(\omega,\beta)
&= & -\omega^4 \int_{\omega^2/\Lambda^{2}}^\infty \frac{dt}{t}\int\frac{d^{4}k}{(2\pi)^{4}}e^{-Bt}\left[ 6+2e^{-t} e^{t\frac{8\beta^2}{B}}\cosh\left( 2t\sqrt{k_\tau^2+\left( \frac{4\beta^2}{B} \right)^2} \right) \right] \nonumber \\ [2mm]
\tilde{V}_{o,F}(\omega,\beta)
&= & 8\omega^4 \int_{\omega^2/\Lambda^{2}}^\infty\frac{dt}{t}\int\frac{d^{4}k}{(2\pi)^{4}}e^{-Bt}
e^{-t/4}\cdot \cosh\left( t\sqrt{k_\tau^2+4\beta^2} \right) 
\end{eqnarray}
where $B=k^2+4\beta^2$.
Since the integral
\begin{eqnarray}
&&-\omega^4 \int_{\omega^2/\Lambda^{2}}^\infty \frac{dt}{t}\int\frac{d^{4}k}{(2\pi)^{4}}e^{-Bt}\left[ 6+2e^{-t} \cosh( 2tk_\tau)-8e^{-t/4}\cosh(tk_\tau)\right]
\nonumber \\[2mm]
\end{eqnarray}
 which is obtained by setting $\beta=0$ in the integrands
 except for the factor $e^{-Bt}$ in the each of contributions vanishes,
 we can subtract it from the total potential.   
Therefore the total potential can be written as
\begin{equation}
\tilde{V}_{o}(\omega,\beta)=\tilde{V}'_{o,B}(\omega,\beta)+\tilde{V}'_{o,F}(\omega,\beta),
\end{equation}
 where
\begin{eqnarray}
\tilde{V}'_{o,B}(\omega,\beta)
&= & -\omega^4 \int_{\omega^2/\Lambda^{2}}^\infty \frac{dt}{t}\int\frac{d^{4}k}{(2\pi)^{4}}e^{-Bt}
2e^{-t} \nonumber \\
&&\times \left[ e^{t\frac{8\beta^2}{B}}\cosh\left( 2t\sqrt{k_\tau^2+\left( \frac{4\beta^2}{B} \right)^2} \right) -
\cosh\left( 2t k_\tau \right) \right], \nonumber \\ [2mm]
\tilde{V}'_{o,F}(\omega,\beta)
&= & 8\omega^4 \int_{\omega^2/\Lambda^{2}}^\infty\frac{dt}{t}\int\frac{d^{4}k}{(2\pi)^{4}}e^{-Bt}
e^{-t/4} \nonumber \\
&& \times \left[ \cosh\left( t\sqrt{k_\tau^2+4\beta^2} \right)  -\cosh\left(t k_\tau\right) 
\right].
\label{subtract}
\end{eqnarray}
We now expand the square brackets in each of the above equations with respect to $t$
 and pick up the terms proportional to $\beta^2$.
We find 
\begin{eqnarray}
\left.e^{t\frac{8\beta^2}{B}}\sum_{n=0}^\infty \frac{4^n}{(2n)!}t^{2n}\left(k_\tau^2+\frac{16\beta^4}{B^2}\right)^n\right|_{\beta^2}&=&8\frac{\beta^2}{k^2}\sum_{n=0}^\infty \frac{4^n}{(2n)!}t^{2n+1}k_\tau^{2n}, \nonumber \\[2mm]
\left.\sum_{n=0}^\infty \frac{1}{(2n)!}t^{2n}\left(k_\tau^2+4\beta^2\right)^n\right|_{\beta^2}&=&4\beta^2\sum_{n=0}^\infty \frac{n+1}{(2n+2)!}t^{2n+2}k_\tau^{2n}.
\end{eqnarray}
Then, by rescaling the integration variable $k$ as
\begin{eqnarray}
\int \frac{d^4k}{(2\pi)^4}\frac{e^{-tk^2}k_\tau^{2n}}{k^2}&=&t^{-(n+1)}\int \frac{d^4k}{(2\pi)^4}\frac{e^{-k^2}k_\tau^{2n}}{k^2},\nonumber \\[2mm]
\int \frac{d^4k}{(2\pi)^4}e^{-tk^2}k_\tau^{2n}&=&t^{-(n+2)}\int \frac{d^4k}{(2\pi)^4}e^{-k^2}k_\tau^{2n},\end{eqnarray} 
 the leading order terms in (\ref{subtract}) become
\begin{eqnarray}
\tilde{V}'_{o,B}(\omega,\beta)
&= & -16\beta^2\omega^4\sum_{n=0}^\infty \frac{4^n}{(2n)!} \int_{\omega^2/\Lambda^{2}}^\infty \frac{dt}{t}e^{-t}t^n\int\frac{d^{4}k}{(2\pi)^{4}}\frac{e^{-k^2}k_\tau^{2n}}{k^2}+\mathcal{O}(\beta^4),\nonumber \\ [2mm]
\tilde{V}'_{o,F}(\omega,\beta)
&= & 16\beta^2\omega^4\sum_{n=0}^\infty \frac{1}{(2n+1)!} \int_{\omega^2/\Lambda^{2}}^\infty \frac{dt}{t}e^{-t/4}t^n\int\frac{d^{4}k}{(2\pi)^{4}}e^{-k^2}k_\tau^{2n}+\mathcal{O}(\beta^4).
\nonumber \\ 
\end{eqnarray}
We can perform the momentum integrations as
\begin{eqnarray}
\int\frac{d^{4}k}{(2\pi)^{4}}\frac{e^{-k^2}k_\tau^{2n}}{k^2}&=&\frac{1}{16\pi^2}\cdot \frac{\Gamma \left(n+\frac12\right)}{\sqrt{\pi}(n+1)},\nonumber \\ [2mm]
\int\frac{d^{4}k}{(2\pi)^{4}}e^{-k^2}k_\tau^{2n}&=&\frac{1}{16\pi^2}\cdot \frac{\Gamma \left(n+\frac12\right)}{\sqrt{\pi}}.
\end{eqnarray}
We find that the summation can be performed as follows: for $\tilde{V}'_{o,B}(\omega,\beta)$, 
\begin{eqnarray}
& & \sum_{n=0}^\infty \frac{4^n}{(2n)!} \int_{\omega^2/\Lambda^{2}}^\infty \frac{dt}{t}e^{-t}t^n\int\frac{d^{4}k}{(2\pi)^{4}}\frac{e^{-k^2}k_\tau^{2n}}{k^2} \nonumber \\
&=& \frac1{16\pi^2}\int_{\omega^2/\Lambda^2}^\infty\frac{dt}{t}e^{-t}\sum_{n=0}^\infty\frac{4^n}{(2n)!}\frac{\Gamma(n+\frac12)}{\sqrt{\pi}}t^n \nonumber \\
&=& \frac1{16\pi^2}\int_{\omega^2/\Lambda^2}^\infty\frac{dt}{t}e^{-t}\sum_{n=0}^\infty\frac{4^n}{(2n)!}\frac{(2n)!}{4^n(n+1)!}t^n \nonumber \\
&=& \frac1{16\pi^2}\int_{\omega^2/\Lambda^2}^\infty\frac{dt}{t}e^{-t}\frac{e^t-1}{t} \nonumber \\
&=& \frac1{16\pi^2}\frac{\Lambda^2}{\omega^2}\left( 1-E_2(\omega^2/\Lambda^2) \right), 
\end{eqnarray}
and for $\tilde{V}'_{o,F}(\omega,\beta)$, 
\begin{eqnarray}
& & \sum_{n=0}^\infty \frac{1}{(2n+1)!} \int_{\omega^2/\Lambda^{2}}^\infty \frac{dt}{t}e^{-t/4}t^n\int\frac{d^{4}k}{(2\pi)^{4}}e^{-k^2}k_\tau^{2n} \nonumber \\
&=& \frac1{16\pi^2}\int_{\omega^2/\Lambda^2}^\infty\frac{dt}{t}e^{-t/4}\sum_{n=0}^\infty\frac1{(2n+1)!}\frac{\Gamma(n+\frac12)}{\sqrt{\pi}}t^n \nonumber \\
&=& \frac1{16\pi^2}\int_{\omega^2/\Lambda^2}^\infty\frac{dt}{t}e^{-t/4}\sum_{n=0}^\infty\frac{2\sqrt{\pi}}{4^{n+1}n!\,\Gamma(n+\frac32)}\frac{\Gamma(n+\frac12)}{\sqrt{\pi}}t^n \nonumber \\
&=& \frac1{16\pi^2}\int_{\omega^2/\Lambda^2}^\infty\frac{dt}{t}e^{-t/4}\cdot\frac12\sum_{n=0}^\infty\frac{\Gamma(n+\frac12)}{\Gamma(n+\frac32)}\frac1{n!}\left( \frac t4 \right)^n \nonumber \\
&=& \frac1{16\pi^2}\int_{\omega^2/\Lambda^2}^\infty\frac{dt}{t}e^{-t/4}F({\textstyle \frac12,\frac32;\frac t4}). 
\end{eqnarray}
Therefore, the total potential becomes 
\begin{eqnarray}
\tilde{V}_o 
&=& \frac{\beta^2\omega^4}{\pi^2}\left( -\frac{\Lambda^2}{\omega^2}\left( 1-E_2(\omega^2/\Lambda^2) \right)+\int_{\omega^2/\Lambda^2}^\infty\frac{dt}{t}e^{-t/4}F({\textstyle \frac12,\frac32;\frac t4}) \right)+{\cal O}(\beta^4). \nonumber \\
\end{eqnarray}
If we also assume $\omega\ll\Lambda$, then 
\begin{eqnarray}
-\frac{\Lambda^2}{\omega^2}\left( 1-E_2(\omega^2/\Lambda^2) \right) 
&=& \log\frac{\omega^2}{\Lambda^2}-1+\gamma+{\cal O}(\omega^2/\Lambda^2), 
\end{eqnarray}
and 
\begin{eqnarray}
\int_{\omega^2/\Lambda^2}^\infty\frac{dt}{t}e^{-t/4}F({\textstyle \frac12,\frac32;\frac t4}) 
&=& E_1(\omega^2/4\Lambda^2)+\int_0^\infty\frac{dt}{t}e^{-t/4}\left( F({\textstyle \frac12,\frac32;\frac t4})-1 \right) \nonumber \\
& & +\int_0^{\omega^2/\Lambda^2}\frac{dt}{t}e^{-t/4}\left( F({\textstyle \frac12,\frac32;\frac t4})-1 \right) \nonumber \\
&=& -\gamma-\log\frac{\omega^2}{4\Lambda^2}+2-\log4+{\cal O}(\omega^2/\Lambda^2)
\end{eqnarray}
imply 
\begin{equation}
\tilde{V}_{o}
=\frac{\omega^4 \beta^2}{\pi^2} =\frac{\omega^2 r^2}{\pi^2} 
\end{equation}
 in the $r<\omega < \Lambda$ region.
The potential in eq.(\ref{small r}) is obtained by analytical continuation back to the Lorentzian signature.



\end{document}